\documentclass[12pt,preprint]{aastex}
\newcommand{\etal}{et~al.}
\newcommand{\CIVdblt}{{\rm C}\kern 0.1em{\sc iv}~$\lambda\lambda 1548, 1550$}
\newcommand{\MgIIdblt}{{\rm Mg}\kern 0.1em{\sc ii}~$\lambda\lambda 2796, 2803$}
\newcommand{\SiIVdblt}{{\rm Si}\kern 0.1em{\sc iv}~$\lambda\lambda 1393, 1402$} 
\newcommand{\CII}{\hbox{{\rm C}\kern 0.1em{\sc ii}}}
\newcommand{\CIII}{\hbox{{\rm C}\kern 0.1em{\sc iii}}}
\newcommand{\CIV}{\hbox{{\rm C}\kern 0.1em{\sc iv}}}
\newcommand{\HI}{\hbox{{\rm H}\kern 0.1em{\sc i}}}
\newcommand{\HII}{\hbox{{\rm H}\kern 0.1em{\sc ii}}}
\newcommand{\Lya}{\hbox{{\rm Ly}\kern 0.1em$\alpha$}}
\newcommand{\Lyb}{\hbox{{\rm Ly}\kern 0.1em$\beta$}}
\newcommand{\Lyg}{\hbox{{\rm Ly}\kern 0.1em$\gamma$}}
\newcommand{\Lyd}{\hbox{{\rm Ly}\kern 0.1em$\delta$}}
\newcommand{\FeII}{\hbox{{\rm Fe}\kern 0.1em{\sc ii}}}
\newcommand{\MgI}{\hbox{{\rm Mg}\kern 0.1em{\sc i}}}
\newcommand{\MgII}{\hbox{{\rm Mg}\kern 0.1em{\sc ii}}}
\newcommand{\OVI}{\hbox{{\rm O}\kern 0.1em{\sc vi}}}
\newcommand{\SiII}{\hbox{{\rm Si}\kern 0.1em{\sc ii}}}
\newcommand{\SiIII}{\hbox{{\rm Si}\kern 0.1em{\sc iii}}}
\newcommand{\SiIV}{\hbox{{\rm Si}\kern 0.1em{\sc iv}}}
\newcommand{\kms}{\hbox{km~s$^{-1}$}}
\newcommand{\cmsq}{\hbox{cm$^{-2}$}}
\newcommand{\cc}{\hbox{cm$^{-3}$}}

\tighten
\begin{document}
 
\accepted{4 10 2001}
\slugcomment{Accepted by {\it The Astrophysical Journal}}
 
\shortauthors{RIGBY ET~AL.}
\shorttitle{WEAK {\MgII} ABSORBERS}


\title{The Population of Weak  {\MgII} Absorbers.  II.  The Properties
of Single--Cloud Systems\altaffilmark{1,2}}

\author{Jane~R.~Rigby,}
\affil{Steward Observatory \\
University of Arizona \\
933 North Cherry Ave. \\
Tucson, AZ 85721 \\
{\it jrigby@as.arizona.edu}}

\vskip 0.15in

\author{Jane~C.~Charlton\altaffilmark{3} and Christopher~W.~Churchill\altaffilmark{4}}
\affil{Department  of  Astronomy  and Astrophysics \\
 The  Pennsylvania State University \\
 University Park, PA 16802 \\
{\it charlton, cwc@astro.psu.edu}}

\altaffiltext{1}{Based  in  part   on  observations  obtained  at  the
W.~M. Keck Observatory, which  is operated as a scientific partnership
among Caltech, the University of California, and NASA. The Observatory
was made possible by the  generous financial support of the W.~M. Keck
Foundation.}
\altaffiltext{2}{Based  in  part  on  observations obtained  with  the
NASA/ESA {\it Hubble Space Telescope},  which is operated by the STScI
for the  Association of Universities for Research  in Astronomy, Inc.,
under NASA contract NAS5--26555.}
\altaffiltext{3}{Center for Gravitational Physics and Geometry}
\altaffiltext{4}{Visiting Astronomer at the W.~M. Keck Observatory}

\begin{abstract}

We present an investigation of {\MgII} absorbers characterized as
single--cloud ``weak systems'' (defined by $W_{r}(2796)<0.3$~{\AA})
at $z\sim 1$.
We measured column densities and Doppler parameters for
{\MgII} and {\FeII} in $15$ systems found in HIRES/Keck spectra at
$6.6$~{\kms}.  Using these quantities and {\CIV}, {\Lya} and Lyman
limit absorption observed with the Faint Object Spectrograph on the
{\it Hubble Space Telescope\/} (resolution $\sim 230$~{\kms}) we
applied photoionization models to each system to constrain
metallicities, densities, ionization conditions, and sizes.

We find that:

1) Single--cloud weak systems are optically thin in neutral hydrogen
and may have their origins in a population of objects distinct from the
optically thick strong {\MgII} absorbers, which are associated with bright galaxies.

2) Weak systems account for somewhere between 25\% to 100\% of the 
$z < 1$ {\Lya} forest clouds in the range $10^{15.8} \leq N({\HI}) \leq
10^{16.8}$~{\cmsq}.

3) At least seven of 15 systems have two or more ionization phases of
gas (multiphase medium).  The first is the low--ionization,
kinematically simple {\MgII} phase and the second is a
high--ionization {\CIV} phase which is usually either kinematically
broadened or composed of multiple clouds spread over several tens of {\kms}.
This higher ionization phase gives rise to the majority of the {\Lya}
absorption strength (equivalent width), though it often accounts for a minor
fraction of a system's $N({\HI})$.

4) We identify a subset of weak {\MgII} absorber, those with
 $\log N({\FeII})/N({\MgII}) > -0.3$, which we term ``iron--rich''.  Though
there are only three of these objects in our sample, their properties are the
best constrained because of their relatively strong {\FeII} detections and the 
sensitivity of the $N({\FeII})/N({\MgII})$ ratio.  These clouds are not
$\alpha$--group enhanced and are constrained to have sizes of $\sim
10$~pc.  At that size, to produce the observed redshift path density, they would
need to outnumber $L^{\ast}$ galaxies by approximately six orders of
magnitude.  The clouds with undetected iron do not have well--constrained 
sizes; we cannot infer whether they are enhanced in their $\alpha$--process 
elements.

We discuss these results and the implications that the weak {\MgII}
systems with detected iron absorption require enrichment from Type Ia
supernovae.  Further, we address how star clusters or supernova
remnants in dwarf galaxies might give rise to absorbers with the
inferred properties. This would imply far larger numbers of such
objects than are presently known, even locally.  We
compare the weak systems to the weak kinematic subsystems in strong
{\MgII} absorbers and to Galactic high velocity clouds.  Though weak
systems could be high velocity clouds in small galaxy
groups, their neutral hydrogen column densities are insufficient for them
to be direct analogues of the Galactic high velocity clouds.

\end{abstract}

%


\section{Introduction}
\label{sec:intro}

Absorption lines from intervening galaxies in quasar (QSO) spectra
provide a wealth of information about the physical conditions of gas
in these galaxies.  Since QSO absorption line spectroscopy offers
unmatched sensitivity to high redshifts and low column densities, the
technique can be used to follow the gas phase in galaxies over cosmic
time, from primordial galaxies to the local universe.

The resonant {\MgIIdblt} doublet has been used extensively to find low
ionization QSO absorption systems (e.g.\ \citet{ltw87} and \citet{ss92}).
At $z\sim1$, {\MgII} absorbers with rest frame equivalent widths,
$W_{r}(2796)$, greater than $0.3$~{\AA} are optically thick in
neutral hydrogen; they are observed to give rise to Lyman limit breaks
\citep{archiveI}.  Furthermore, these ``strong'' systems almost
always arise within $40 h^{-1}$~kpc of bright ($L \geq 0.05~L^{\ast}$), normal
galaxies \citep{bb91,sdp94,steidel95}.  The gas
kinematics are consistent with material in the disks and extended
halos of galaxies \citep{lb92,pb90,csv96,jcccwc,datapaper}.

The discovery of weak systems (those with $W_{r}(2796) < 0.3$~{\AA})
in high resolution HIRES/Keck spectra (\citet{weak}; hereafter Paper I)
necessitates a revision in the standard picture.  Weak systems
comprise $\sim 65$\% of {\MgII} selected absorbers by number, yet only
4 out of 19 whose fields have been imaged have a $\ga 0.05~L^*$ galaxy
candidates within $\simeq 50 h^{-1}$~kpc\footnote{More precisely, fields were
searched within $10 \arcsec$ of the quasar, which corresponds to
$\simeq 50 h^{-1}$~kpc at $z\sim1$} \citep{chuckprivcomm}.  
The large redshift path
number density of weak systems relative to that of Lyman limit systems
(LLSs) statistically indicates that, unlike strong absorbers, the
majority of weak systems arise in optically thin neutral hydrogen
(sub--Lyman limit) environments (see Figure 5 of
Paper~I\nocite{weak}).  This has been observationally confirmed by
\citet{archiveI}.  Furthermore, the majority
of weak systems are single clouds, often unresolved in HIRES/Keck
spectra (resolution $6.6$~{\kms}), in striking contrast to the complex
kinematics of strong {\MgII} absorbers \citep{pb90,datapaper}.
This evidence suggests that a substantial fraction
of the weak {\MgII} systems selects a population of objects distinct
from the bright, normal galaxies that are selected by the strong
{\MgII} absorbers.

This begs the question: what are these single--cloud objects that
outnumber {\MgII}--absorbing galaxies, yet often have no obvious
luminous counterparts?  A first strategy for addressing this question,
adopted in this paper, is to constrain the column densities, 
metallicities, ionization conditions, and sizes of single--cloud weak systems.

Setting useful constraints on these physical conditions requires not
just low ionization species, but also neutral hydrogen and medium
to high ionization species.  At $z\sim1$, the Lyman series and the
{\CII}, {\CIV}, and {\SiIV} transitions fall in the near
ultraviolet (UV).  These transitions were observed with the
low--resolution Faint Object Spectrograph onboard the {\it Hubble
Space Telescope} (FOS/{\it HST}).  \citet{1206}
have demonstrated that photoionization modeling
using both high--resolution optical spectra and low--resolution UV
spectra can yield meaningful constraints on the physical conditions of
{\MgII} absorbers.  We have adapted their approach to study
weak systems.

In \S~\ref{sec:data} we describe the optical and UV spectra used in
this study, and discuss sample selection to motivate our focus on
single--cloud weak {\MgII} systems.  In
\S~\ref{sec:models}, we describe our methods for Cloudy \citep{ferland}
photoionization modeling to obtain constraints on cloud metallicities,
ionization conditions, and sizes.  The resulting constraints for the
fifteen single--cloud weak {\MgII} systems are presented in
\S~\ref{sec:cloudnotes} and summarized in \S~\ref{sec:sumcloud}.  In
\S~\ref{sec:discussion} we consider what types of astrophysical
environments could be consistent with the properties of the weak
single--cloud {\MgII} absorbers.  In
\S~\ref{sec:further} we speculate about the evolution of weak {\MgII}
systems and suggest future investigations.


\section{Data and Sample}
\label{sec:data}

\subsection{The HIRES Spectra}
\label{sec:hires}

Weak systems were charted in a survey of $26$ QSO spectra
(Paper~I\nocite{weak}) obtained with the HIRES spectrograph
\citep{vogt94} on the Keck I telescope.  A total of 30 systems were
found, with 22 of them being new discoveries.  The survey covered the
redshift interval $0.4 \leq z \leq 1.4$, and was unbiased for
$W_{r}(2796)<0.3$~{\AA}.  The spectral resolution was $R=45,000$ ({\sc
fwhm} = 6.6$~{\kms}$) with a typical signal--to--noise ratio of $S/N
\simeq 30$ per three--pixel resolution element.  The survey was $80$\%
complete for a $5\sigma$ equivalent width detection threshold of
$W_{r}(2796)=0.02$~{\AA}.  We restrict our study to those systems with
this limiting equivalent width in the continuum at the $\lambda 2796$
transition.  Simulations reveal that Voigt profile fits to HIRES data
become increasing less certain below this limit \citep{thesis,datapaper}.
This equivalent width cutoff removes from this sample three systems
from Paper I: S9, S11, and S22.

Data reduction, line identification, and {\MgII} doublet identification
have been described in Paper~I\nocite{weak}.  In Figure
\ref{fig:spectra}  we  display all the detected transitions ({\MgII},
{\FeII}, and {\MgI}) associated with the $16$ weak single--cloud
{\MgII} absorbers found in regions of the HIRES spectra that satisfy
our equivalent width selection criterion.

\subsection{Physically Motivated Sample}

The adopted equivalent width demarcation at
$W_{r}(2796)=0.3$~{\AA} between ``weak'' and ``strong'' {\MgII}
absorbers is an artifact of observational sensitivity (e.g.\
\citet{ss92}).  However, there are at least two physical
conditions that set weak systems apart from strong systems.  First,
almost all weak systems are optically thin to neutral hydrogen.  This
was shown statistically in Paper I and observationally in
\citet{archiveI}.  By contrast, strong {\MgII}
absorbers are Lyman limit systems, and thus by definition are optically
thick in neutral hydrogen.  

Second, the weak systems are often single clouds with very small 
velocity widths (unresolved in the HIRES spectra).
For the sample of $30$ weak systems presented in Paper I, the number
of clouds per weak system ranges from $1$ to $7$.  
Figure~\ref{fig:nclouds} shows the frequency
distribution of the number of clouds per system for both the strong
systems and the weak systems.  The number of clouds per strong
absorption system follows a Poissonian distribution with a median of $7$ clouds
(see Table~7 of \citet{datapaper}).  In contrast, the
distribution for weak systems is
non--Poissonian, with a spike at one cloud per absorber, i.e.\
there are many single--cloud weak systems.  This suggests
that these {\MgII} absorbers represent a distinct population of
objects.  To produce many more single--cloud absorbers than
multiple--cloud absorbers, weak systems should have a small covering
factor or a preferred large--scale geometry, such that a line of sight
is unlikely to intersect multiple clouds.

In this paper we study the single--cloud systems.  This selection
criterion, together with the equivalent width cutoff described in 
\S~\ref{sec:hires}, yield $16$ single--cloud systems from the sample of 
$30$ systems of Paper I\nocite{weak}.

\subsection{The FOS Spectra}

For $13$ of the $16$ single--cloud systems, archival FOS/{\it HST\/}
spectra were available.  The resolution of these spectra was
$\sim 230$~{\kms}~{\sc fwhm},
with four diodes per resolution element.  For PKS $0454+039$ and PKS
$0823-223$, which contain three single--cloud weak systems, the
spectra were retrieved from the archive and reduced in collaboration
with S.  Kirhakos, B. Jannuzi, and D.  Schneider \citep{archiveI}, 
using the techniques
and software of the {\it HST\/} QSO Absorption Line Key Project
\citep{KP-dps}.  The remaining six QSOs were observed and
published by the Key Project \citep{cat1,cat2,cat3}
and have kindly been made available for this work.  Further details of
the FOS/{\it HST\/} observations and the analysis were presented
in \citet{archiveI}.  In general, for the
single--cloud weak systems discussed here, the useful FOS
transitions were {\CIV}, {\Lya}, and the Lyman limit, since they are
strong features commonly covered in the archive data.
Table~\ref{tab:tab1} lists the equivalent widths and limits for
{\CIV} and {\Lya}, taken from \citet{archiveI}\footnote{Weak systems in Paper~I were
numbered in increasing redshift order; we have adopted those system
numbers.}.

In Figure~\ref{fig:array}, spectra covering {\MgII}, {\FeII}, {\CIV},
{\Lya} and the Lyman limit are displayed for each system to
show, in a single view, the full sample including blends,
non--detections, and non--coverage.  Additional transitions observed
in the FOS data but not plotted in Figure~\ref{fig:array} can be
found in \citet{archiveI}.

\subsection{Column Densities}
\label{sec:spec_vp}

For the HIRES spectra, the column densities and Doppler parameters
were obtained using {\sc Minfit}, a $\chi^{2}$ minimization Voigt
profile fitter \citep{thesis,datapaper,cvc01}.
The column densities and Doppler
parameters of the fits to {\MgII} and {\FeII} are listed in
Table~\ref{tab:tab1}, as are $3\sigma$ upper limits on the
{\FeII} column density for cases where no {\FeII} transitions
were detection.  The latter were obtained from the $3\sigma$
equivalent width limit for {\FeII} $\lambda 2600$, assuming
a Doppler parameter equal to that obtained for {\MgII} in the
same system\footnote{Limits were insensitive to the assumed Doppler
parameter because {\FeII} was on the linear part of the curve of
growth.}.

The column density ratio, $N({\FeII})/N({\MgII})$, is critically
important to the photionization modeling; as discussed in 
\S~\ref{sec:constraints}, this ratio constrains the ionization parameter.
Therefore, it is important to appreciate the systematic errors when 
$N({\MgII})$ and $N({\FeII})$ are comparable in strength.  Simulations 
reveal that a $5\sigma$ equivalent width detection limit of $0.02$~{\AA} 
in HIRES spectra gives a 99\% completeness limit of 
$\log N({\MgII}) = 11.9$~{\cmsq} and $\log N({\FeII}) = 12.4$~{\cmsq}
\citep{thesis,cvc01}.  Above these column densities {\sc Minfit} accurately
models the column densities.  For $\log N({\FeII})<12.4$~{\cmsq}, {\sc
Minfit} statistically overestimates $\log N({\FeII})$ by up to
$\sim0.3$~dex (\citet{thesis}, Figure~4.8) due to a bias toward ``false detections''
in ``favorable'' noise patterns.  {\MgII} Doppler parameters
were well recovered in the simulations, with an {\sc RMS} scatter of
$\sim 1$~{\kms} for all column densities above the sensitivity cutoff. 
{\FeII} Doppler parameters were poorly recovered when the column
density was near or below the sensitivity cutoff \citep{thesis}.

For an {\FeII} detection at the $3\sigma$ level in the lowest
$S/N$ spectrum in our sample, the detection limit (99\% completeness) for the ratio
$N({\FeII})/N({\MgII})$ would be
\begin{equation}
\log  [N({\FeII})/N({\MgII})] = 12.2 - \log N({\MgII}) ,
\label{eq:fe2mg2}
\end{equation}
Therefore, with  increasing $N({\MgII})$ the upper limit  
on   $N({\FeII})/N({\MgII})$ decreases, i.e.\ becomes more stringent.  
We show this relation in Figure~\ref{fig:femg-grid} as a diagonal
line.  The value of $\log N({\FeII})/N({\MgII})$ for each of the
single--cloud weak systems is plotted, with detections as solid circles and
upper limits as downward--pointing arrows.  Note that since the 
plotted $N({\FeII})/N({\MgII})$ detection limit line was computed
from the noisiest spectrum in our sample, the data points are often lower
than the limit, and thus are more constraining of the {\FeII} to {\MgII}
ratio than the line indicates.

Since there is an apparent gap in the distribution of 
$\log N({\FeII})/N({\MgII})$,
we somewhat arbitrarily define systems with detectable {\FeII} and 
$\log N({\FeII})/N({\MgII}) > -0.3$ (systems S7, S13, and S18) 
as ``iron--rich'' weak systems.

Since all three so--called ``iron--rich'' systems lie above the 
worst--case sensitivity line, and all other systems fall below it,
there may be a selection effect at work -- are the highest $N({\MgII})$ systems
identified as ``iron--rich'' systems simply because their {\FeII}
is easier to detect?  We argue that this is not the case.
To illustrate, we consider the five systems with 
$N({\MgII})$ between that of S18 and S7 (two iron--rich systems.)  
If these five systems had 
$\log N({\FeII})/N({\MgII})\sim~0$, as is true for the iron--rich systems,
then Figure~\ref{fig:femg-grid} makes clear that they should have
detected {\FeII}.  Only one of the five systems does (S28), and since it 
has $\log N({\FeII})/N({\MgII}) \simeq -0.6$, it is not deemed 
iron--rich.  The limits on $\log N({\FeII})/N({\MgII})$ for the 
other four systems are also well below $0$, by more than 
half a dex.  Thus, it seems that while
we have insufficient information to address whether the distribution of 
$N({\FeII})/N({\MgII})$ is continuous or bimodal, the iron--rich 
systems do appear to have significantly higher $N({\FeII})$ to $N({\MgII})$
ratios than do the other systems.  As \S~\ref{sec:constraints} will show,
this difference in column density ratios indicates variations in 
ionization and/or abundance pattern between the ``iron--rich'' systems and 
the other systems.

\subsection{Photoionization Modeling}
\label{sec:models}

We assume that the single--cloud, weak systems are in photoionization
equilibrium and constrain their properties with the photoionization
code Cloudy, version 90.4 \citep{ferland}.  The clouds are modeled as
constant density, plane--parallel slabs and are matched to the {\MgII}
and {\FeII} column densities measured from the HIRES spectra.  Using
each model's output column densities for ions of interest, we synthesized
FOS/{\it HST\/} spectra, which we directly compared to the observed
{\CIV} and {\Lya} profiles and to the spectral region covering
the Lyman limit break.  This procedure constrains both the ionization
conditions and gas--phase metallicities.  Further discussion of the
modeling technique was presented by
\citet{1206} and by \citet{q1634}.

We begin each analysis with the assumption that the {\MgII}, {\FeII},
{\Lya}, and {\CIV} arise in a single isothermal structure, described
by a single metallicity and ionization parameter.  The data often
show that this assumption is violated, in which case we model two phases,
each having its own metallicity, temperature, and ionization parameter.
Further details are presented below.

Clouds were assumed to be ionized by a Haardt \& Madau background
spectrum \citep{haardtmadau96}.  For cloud redshifts below $z=0.75$,
we used a spectrum shape and normalization at $z=0.5$ and for
redshifts above $z=0.75$ we used the shape and normalization at
$z=1.0$.  For all models, we assumed a solar abundance pattern.  In
\S~\ref{sec:robust}, we discuss  possible abundance pattern variations
and the effect of alternative spectral shapes.

\subsubsection{Applying Constraints to the {\MgII} Cloud}
\label{sec:constraints}

As Figure \ref{fig:ratios} illustrates, when the assumption of photoionization
equilibrium holds, the $N({\FeII})/N({\MgII})$ and $N({\CIV})/N({\MgII})$ ratios
are uniquely determined by the ionization parameter, which is defined
as $U = n_{\gamma}/n_{\rm H}$, where $n_{\gamma}$ and $n_{\rm H}$ are
the number density of photons capable of ionizing hydrogen and the
total hydrogen number density, respectively.  Therefore, 
$\log U = \log n_{\gamma} - \log n_{\rm H}$, where $n_{\gamma}$ is set by the
background spectrum.  For the
\citet{haardtmadau96} spectrum, $\log n_{\gamma}=-5.6$ at
$z=0.5$ and $\log n_{\gamma}=-5.2$ at $z=1.0$.

When $N({\MgII})$ and $N({\FeII})$ were both measured, they were used
to constrain the optimized mode of Cloudy at a set metallicity to
yield $U$ and $N({\HI})$ by varying both $N({\HI})$ and $n_{\rm H}$.
Models were run for a range of metallicities, $-2.5\leq Z\leq0$, in
increments of $0.5$ dex\footnote{We use the notation $Z \equiv \log
(Z/Z_{\odot})$.}.  

When only an upper limit on $N({\FeII})$ was
measured, we created a grid of Cloudy models over $\log U$ (from $-2$
to $-5$, in $0.1$ dex intervals) and $Z$ (from $0$ to $-2.5$, in $0.5$
dex intervals.)  $N({\MgII})$ provided the constraint, and $N({\HI})$
was the parameter for which Cloudy solved.  We rejected ionization
parameters whose models, over the whole metallicity range, yielded
$N({\FeII})$ that was greater than the $3\sigma$ limits listed in Table 1.
This set lower limits on the ionization parameter.

A model was judged to have failed or be inapplicable for any of
the following three reasons: failure to converge upon an ionization
parameter (usually when $N({\FeII})/N({\MgII}) \simeq 1$); cloud size
exceeded the Jeans' length and was thus unstable to collapse; cloud
size exceeded $50$~kpc.

\subsubsection{Metallicities and Multiple Ionization Phases}
\label{sec:metallicities}

Whereas the ionization parameter is constrained by the {\MgII} 
and {\FeII} HIRES data, the metallicity and $N({\HI})$ are 
constrained by the {\Lya} profiles and presence or absence 
of the Lyman limit break in the FOS spectra.
In the regime where a cloud is optically
thin at the Lyman limit, to create more $W_r({\Lya})$, given a
measured $N({\MgII})$, requires lowering the metallicity, which
increases $N({\HI})$ the total hydrogen column density
$N_{\rm H}$, where $N_{\rm H}=N({\HI}) + N({\HII})$, and
the cloud size.

Because the low-resolution FOS/{\it HST\/} profiles are largely
dominated by the instrumental spread function, their column densities
and Doppler parameters cannot be directly measured.  In order to use
the FOS spectra as a constraint, we created synthetic FOS spectra
(infinite sampling and signal--to--noise ratio) using the kinetic
temperature and column densities output by Cloudy.

First, we assumed {\it a priori\/} that all detected absorption arises
in one phase of gas, the same phase that gives rise to {\MgII}
absorption.  Using the temperature of the Cloudy model and the
measured $b({\MgII})$, we solved for the turbulent component,
$b_{turb}$, and computed the Doppler parameter for other ions by the
relation
\begin{equation}
b_{ion}^{2} = 2kT/m_{ion} + b_{turb}^{2} .  
\label{eq:b}
\end{equation}
This inferred Doppler parameter and the column density are used to
generate a Voigt profile, which is then convolved with the FOS
instrumental spread function to produce a synthesized spectral feature.

To constrain each cloud's metallicity, we visually compared the
synthesized and FOS {\Lya} profiles for each modeled metallicity.  In
Figure~\ref{fig:metdemo}, we illustrate the metallicity fitting
procedure.  Metallicities of $Z=0$, $-1$, $-2$, and $-2.5$ are
plotted; clearly, $Z=0$ and $Z=-1$ do not fit the {\Lya} profile nor
are their equivalent widths consistent with the measured value,
and $Z=-2.5$ slightly overpredicts the {\Lya} absorption.  In
this example, assuming a single phase produces the {\Lya}
absorption, the metallicity is constrained to be in the range
$-2.5<Z<-1$.

For the systems for which the Lyman limit break is known to be absent, an
additional metallicity constraint may be imposed.  From the {\Lya}
curve of growth, shown in Figure~\ref{fig:cog}, for a given $N({\HI})$
we can find the $b({\HI})$ required to observe a
given $W({\Lya})$.  The lack of a Lyman limit break provides a direct
constraint, $\log N({\HI}) < 16.8$~{\cmsq}, which then implies a lower
limit on $b({\HI})$.  If this $b({\HI})$ is larger than that implied
by thermal scaling of $b({\MgII})$ for the cloud, then there is
evidence for a second phase of gas.  Much of the {\Lya} equivalent
width must arise in this second phase, whose larger Doppler parameter
produces the observed $W_r({\Lya})$ without exceeding the limit on
$N({\HI})$ (imposed by the Lyman limit).  With such a second
phase present, the $N({\HI})$ would be smaller in the {\MgII} cloud phase,
and the metallicity ``constraint'' should be taken as a lower limit.

The strength of the {\Lya} profile depends only weakly on the
ionization parameter for most of the range under consideration.
Accordingly, when the ionization parameter was poorly constrained, we
constrained the metallicity at the low and high ends of the permitted
$U$ range, so that the quoted metallicity constraint holds for the
range of possible ionization parameters.

\subsubsection{Multiple Ionization Phases Required by {\CIV}}
\label{sec:multc4}

Often, the single--phase {\MgII} cloud model could not account for the
observed {\CIV} strength, even for the highest possible
ionization parameter.  If {\FeII} is detected, it imposes a strict upper
limit on the ionization parameter, and therefore on $N({\CIV})$.  If
{\FeII} is not detected, a high ionization parameter and therefore a
large $N({\CIV})$ is possible.  However, with a small $b({\CIV})$,
scaled from the measured $b({\MgII})$, it may still not be possible
for the model to reproduce the observed $W_{r}({\CIV})$.
Single--phase models were rejected when the equivalent widths of the 
synthesized profiles fell at least $3\sigma$ below that measured in the FOS data;
a second phase with a larger $b({\CIV})$ was then required.
This second phase should be sufficiently ionized that
it does not produce a detectable broad {\MgII} component.
However, its other properties (eg. its Doppler parameter
and metallicity) are not well constrained by the low
resolution FOS spectrum.

The {\CIVdblt} doublet ratio can also indicate that {\CIV} absorption
occurs in a broader phase.  If all transitions arose in the narrow
{\MgII} phase, the {\CIV} doublet would be saturated and
unresolved, with a doublet ratio of $\sim1$.  By contrast, a broad
phase would produce less saturated lines, and a {\CIV} doublet ratio
closer to the natural value of $2$.  This argument is not
model--dependent but relies only on the spectral resolution
of FOS/{\it HST\/}.  Unfortunately, the errors in the
doublet ratio are large; S17 is the only system for which the doublet
ratio can be used to infer that the {\CIV} arises in a broader
phase than the {\MgII}.

\subsubsection{Inferred Sizes and Masses}
\label{sec:sizes}

The derived cloud size (plane parallel thickness) is simply 
$d = N_{\rm H}/n_{\rm H}$, where $N_{\rm H}$ is the total hydrogen 
column density, $N({\HI}) + N({\HII})$.  Assuming a spherical geometry,
we can estimate the cloud masses within a geometric factor as
\begin{equation}
M_{cl} \simeq 4~\bigg( \frac{N_{\rm H}}{10^{18}~{\rm cm}^{-2}} \bigg) ^{3} 
        \bigg( \frac{10^{-2}~{\rm cm}^{-3}}{n_{\rm H}} \bigg) ^{2} 
        \quad {\rm M}_{\odot} .
\label{eq:mass}
\end{equation}
Because $n_{\rm H} = n_{\gamma}/U$,  this equation can also be written
in terms  of the ionization  parameter where $n_{\gamma}$  is slightly
redshift dependent.

\subsubsection{Robustness to Assumptions}
\label{sec:robust}

Much of what can be inferred and/or deduced about the physical and
cosmological properties of single--cloud weak
systems is directly related to the size and mass measurements from the
models.  Here, we examine the sensitivity of these quantities to our
modeling assumptions.

First, we have assumed photoionization equilibrium.  Photoionization
models have been shown to underestimate the sizes of some {\Lya}
forest clouds by several orders of magnitude \citep{haehnelt}.  In
these cases, the gas is not in full thermal equilibrium because of
additional heating from shocks, and errors in temperature propagate to
large errors in derived size.  Such non--equilibrium is found to occur
for $\log n_{\rm H} \le -4$~{\cc}, corresponding to the high
ionization conditions of higher redshift {\Lya} forest clouds, where
recombination timescales can rival a Hubble time.  At $z=1$,
such a density corresponds to $\log U = -1.4$.  As shown in
Figure~\ref{fig:ratios}, photoionized clouds with $N({\FeII}) \simeq
N({\MgII})$ are constrained to have $\log U \leq -3.5$.  Thus,
iron--rich weak {\MgII} systems have densities too high,
or equivalently ionization parameters too low, for their sizes
to be underestimated in this manner.

Second, we have assumed that the gas is ionized by a
\citet{haardtmadau96}
extragalactic UV background spectrum.  Since high luminosity
counterparts are apparently rarely associated with 
weak systems, the most likely stellar contribution to the spectrum
would be that from a single star or small group of stars quite near to
the cloud.  However, the constraints on the stellar types, number of
stars and their distance from the cloud can be quite severe
(e.g. \citet{cwc-lb98}, Appendix B).  To explore model
sensitivity to the spectral shape, we produced a Cloudy grid, similar
to that in Figure~\ref{fig:ratios}, for a stellar spectrum
characteristic of $T=30,000$~K stars.  Such a spectrum, though it is
too soft to ionize carbon into {\CIV}, still has {\FeII}/{\MgII} that
decreases with increasing ionization parameter.  Thus a low ionization
parameter (and high density) is still required for the {\MgII}
clouds, especially those with measured {\FeII}.

Third, we have assumed a solar abundance pattern with no depletion
onto dust grains.  Enhancement of $\alpha$--process elements relative
to iron would shift the Cloudy
grid in Figure~\ref{fig:femg-grid} down, so that a
given $N({\FeII})/N({\MgII})$ ratio would correspond to
a lower ionization parameter and smaller cloud size.
By contrast, iron enhancement would make the clouds larger than we infer, but
such enhancement is rarely found in astrophysical environments \citep{edv93,mcwilliam}.
Dust depletion would mimic $\alpha$ enhancement, since iron depletes more
readily than magnesium by as much as 0.5 dex (e.g.\ \citet{laur} and \citet{savage}).
Dust would also affect the cooling function in the models.
To investigate model sensitivity to this, we ran Cloudy models for S7
(a high $N({\FeII})/N({\MgII})$ system) both with no dust and with varying amounts
of dust scaled relative to the ISM level (taken from the Cloudy
database).  Dust had a negligible effect on cloud sizes for dust
abundances up to ten times ISM.  At higher levels, the
cloud sizes decreased with increasing dust content.

To summarize, the model cloud properties are robust to the modeling 
assumptions.  In particular, the conclusion seems robust that clouds with 
$N({\FeII}) \simeq N({\MgII})$ and low $N({\HI})$ are small.  If some unknown
effect has led us to underestimate the cloud sizes of the iron--rich
clouds by $2$--$3$ orders of magnitude, then the discussion of their
origins in \S\ref{sec:highfe} does not apply, and the discussion in
\S\ref{sec:lowfe} for clouds with lower $N({\FeII})/N({\MgII})$ would be more
suitable.


\section{Individual System Properties}
\label{sec:cloudnotes}

For each single--cloud weak system, we first summarize in brackets the coverage
and detection status of {\FeII}, {\Lya}, the Lyman limit, and {\CIV}.
We then explain the constraints these transitions impose.  Lastly, in brackets
we summarize the evidence or lack of evidence for multiple phases of
gas.  Systems are listed in redshift order, with system numbers
adopted from Paper I.  The constraints for all systems are summarized
in Table~\ref{tab:tab2}.

\subsection{\rm S1  (Q$1421+331$;  $z_{\rm abs}=0.45642$)}

[{\FeII}  not covered;  no FOS  spectra.] Since  {\Lya} and  the Lyman
limit  were not  covered, the  metallicity  of this  system cannot  be
constrained.  The  ionization parameter cannot be constrained because
{\FeII} and {\CIV} were not covered.  We do not include this system in
discussions of  multiple phases and {\FeII} statistics  because of the
lack of spectral coverage.  [Cannot address multiphase.]

\subsection{\rm S3  (Q$1354+195$; $z_{\rm abs}=0.52149$)}

[$N({\FeII})$  upper limit;  $W_{r}({\Lya})$ measured;  no break  at Lyman
limit;  $W_{r}({\CIV})$  upper  limit.]   To  produce  the  strong  {\Lya}
detection  in this  cloud  requires $Z<-2.5$,  assuming  that all  the
{\Lya}  absorption arises in  the same  phase of  gas as  the detected
{\MgII}.   However, the  lack of  a Lyman  limit break  requires $\log
N({\HI})<16.8$~{\cmsq}, which corresponds  to $Z>-1.5$ for this cloud.
Therefore, S3  possesses a second phase  of gas with  a larger Doppler
parameter than that of the {\MgII} phase, which can produce the
observed {\Lya}  equivalent width  without exceeding the  {\HI} column
density limit imposed  by the Lyman limit.  {\FeII}  and {\CIV} limits
do not  constrain $U$.  [Multiphase required because  of {\HI}; {\CIV}
does not require or rule out multiphase.]

\subsection{\rm S6  (Q$0002+051$; $z_{\rm abs}=0.59149$)}

[$N({\FeII})$  upper limit; {\Lya}  in spectropolarimetry  mode; Lyman
limit not  covered; $W_{r}({\CIV})$ upper limit.]  The  {\Lya} spectrum is
not   usable  because   it  was   taken  in   spectropolarimetry  mode
\citep{archiveI}.   This,  combined  with  the lack  of  Lyman  limit
coverage, allows no constraints on metallicity.  The {\FeII}
limit  sets a  fairly high  lower limit  on the  ionization parameter:
$\log  U >  -3.5$.   The {\CIV}  limit  is too  poor  to restrict  the
ionization parameter.  [Cannot address multiphase.]

\subsection{\rm S7  (Q$0454+039$; $z_{\rm abs}=0.64283$)}

[$N({\FeII})$ measured; $W_{r}({\Lya})$ measured; Lyman limit not covered;
$W_{r}({\CIV})$ measured.]  If we assume  the {\Lya} arises in the {\MgII}
phase, then  to match the  {\Lya} profile requires  $-1<Z<0$. [This
constraint differs somewhat from \citet{cwc-lb98}
who used a different version of the FOS/{\it HST} spectrum.]
{\MgII} and {\FeII} constrain the ionization parameter to $-4.5< \log U <-3.6$
for  the full range  of metallicities  modeled, and  to $-4.5<  \log U
<-4.2$  for  the  metallicity   range  determined  above.   The  model
thicknesses range from $2$~pc for  $Z=0$ to $8$~pc for $Z=-1$.  If the
cloud were enhanced in iron,  both size and ionization parameter would
increase.  The  large {\FeII}/{\MgII} ratio requires  a low ionization
cloud which  cannot produce the  observed {\CIV} or {\SiIV};  a second
phase of higher--ionization gas is required.  [Multiphase is not required
for {\HI}, but is required to explain {\CIV} and {\SiIV}.]

\subsection{\rm S8  (Q$0823-223$; $z_{\rm abs}=0.705472$)}

[$N({\FeII})$  upper  limit;  {\Lya}  and  Lyman  limit  not  covered;
$W_{r}({\CIV})$  upper limit.]  Neither  {\Lya} nor  the Lyman  limit were
covered, providing no metallicity  constraint.  The {\FeII} and {\CIV}
limits  set lower and  upper limits,  respectively, on  the ionization
parameter: $-3.6< \log U <-2.4$.  In order not to exceed the $3\sigma$
equivalent  width,  $\log  N({\CIV})  <  14$~{\cmsq}.   [{\HI}  cannot
address multiphase; {\CIV} is consistent with a single phase.]

\subsection{\rm S12  (Q$1634+706$; $z_{\rm abs}=0.81816$)}

[$N({\FeII})$ upper  limit; $W_{r}({\Lya})$ poor upper  limit; Lyman limit
not covered; $W_{r}({\CIV})$ upper  limit.]  The available {\Lya} coverage
is pre--COSTAR  and in spectropolarimetry mode,  and therefore unusable
\citep{archiveI}; accordingly, the  metallicity of this system cannot
be  constrained.   The  {\FeII}  limit requires  $\log  U>-4.4$.   The
$3\sigma$ {\CIV} limit, which sets $\log N({\CIV}) < 15$~{\cmsq}, does
not  restrict  $U$.   [{\HI}  cannot  address  multiphase;  {\CIV}  is
consistent with a single phase.]

\subsection{\rm S13  (Q$1421+331$; $z_{\rm abs}=0.84325$)}

[$N({\FeII})$ measured; no FOS spectra.]  Without FOS spectra for this
quasar,      the      metallicity      cannot     be      constrained.
$N({\FeII})/N({\MgII})$ is  greater than  unity for this  cloud: $\log
N({\FeII})        =       13.47\pm0.07$~{\cmsq}        and       $\log
N({\MgII})=13.1\pm0.1$~{\cmsq}.    As   Figures~\ref{fig:femg-grid} and
\ref{fig:ratios} demonstrate,  this   cannot  occur  for  a  solar
abundance pattern  and a Haardt  \& Madau (1996)  ionizing background.
The iron  fit is particularly robust because  five {\FeII} transitions
were   detected  at  relatively   high  $S/N$.    Unresolved
saturation  may be  present in  the {\FeII}  transitions,  which would
cause the  fit to {\it underestimate\/} $N({\FeII})$.   If the {\MgII}
doublet does  not have  unresolved saturation, then  it could  be that
this  cloud is iron enhanced, as depletion cannot yield this pattern.  
A model with iron enhanced by $0.5$~dex predicts
$\log U  = -4.3$  and cloud size  from $1$--$12$~pc.  [Without
FOS spectra, multiphase cannot be addressed.]

\subsection{\rm S15  (Q$0002+051$; $z_{\rm abs}=0.86653$)}

[$N({\FeII})$  upper limit;  $W_{r}({\Lya})$ measured;  no break  at Lyman
limit; $W_{r}({\CIV})$  upper limit.]   If the observed  {\Lya} absorption
were  produced  in  the  {\MgII}  cloud, this  would  require  a  cloud
metallicity of $-2.5 \leq Z \leq -2.0$, which corresponds to $18< \log
N({\HI}) <20$~{\cmsq}.   However, the  small Doppler parameter  of the
{\MgII} cloud  predicts an  unresolved {\Lya} narrower  than observed,
and the  lack of a  break at the  Lyman limit\footnote{The apparent break
in Figure~\ref{fig:array} results entirely from a strong {\MgII} absorber
at $z=0.8514$ \citep{archiveI}.} requires  $\log N({\HI})
\leq 16.8$~{\cmsq}  and $Z>-1$ for all  ionization parameters.  Hence,
much of the {\Lya} absorption arises  not in the {\MgII} cloud, but in
a separate phase of gas  with a larger Doppler parameter
(see Figure~\ref{fig:cog}).  The {\FeII}
limit constrains  $\log U>-3.6$ in  the {\MgII} cloud.  Models  can be
made to  meet but not exceed  the {\CIV} limit; therefore  it does not
constrain  the  ionization  parameter;  $\log  N({\CIV})<16.5$~{\cmsq}
could  exist in  the  {\MgII} cloud.   [Multiphase is inferred from {\HI},
though {\CIV} does not require a second phase.]

\subsection{\rm S16  (Q$1241+176$; $z_{\rm abs}=0.89549$)}

[$N({\FeII})$  upper  limit;  $W_{r}({\Lya})$  measured; Lyman  limit  not
covered;  $W_{r}({\CIV})$  upper  limit.]    If  the  {\Lya}  and  {\MgII}
absorption  arose  in the  same  phase, then  {\Lya} would be best fit  by
$-2.5<Z<-1$ (see Figure~\ref{fig:metdemo}).  
The ionization parameter is constrained as $-4.8< \log U <-2.0$
by {\FeII}  and {\CIV} limits.  The upper  limit of $N({\CIV})$
from     the    $3\sigma$     equivalent     width    limit,     $\log
N({\CIV})<13.4$~{\cmsq}, can be produced in the {\MgII} phase and thus
does not  require a  second phase.  However,  if the Lyman  limit were
covered,   its  metallicity   constraint  might   conflict   with  the
{\Lya}--determined metallicity,  and thus necessitate  a second phase.
[{\HI}  cannot address  multiphase;  {\CIV} is consistent  with a  single
phase.]

\subsection{\rm S17  (Q$1634+706$; $z_{\rm abs}=0.90555$)}

[$N({\FeII})$  upper  limit;  $W_{r}({\Lya})$  measured; Lyman  limit  not
covered; $W_{r}({\CIV})$ measured.]  If  all detected absorption arose in
one phase,  then {\Lya} would best  be fit by  $-2<Z<-1$, though these
fits  do not  match the  observed wings.   However, the  small Doppler
parameter of  the {\MgII} phase predicts  unresolved, saturated {\CIV}
profiles with  a doublet  ratio of 1,  which is inconsistent  with the
observed   doublet   ratio   of   $1.4   \pm   0.2$.
So  while the  $\log  U=-2.0$ models  can
reproduce   the  {\CIV}   $\lambda  1551$   equivalent   width  ($\log
N({\CIV})\simeq  17$~{\cmsq}), they  cannot fit  both  {\CIV} $\lambda
1548$ and {\CIV} $\lambda 1551$ at  once.  A second phase of gas, with
a larger effective Doppler parameter,  is required to give rise to the
less  saturated  {\CIV} profile.   The  {\FeII}  limit constrains  the
ionization  parameter in the  {\MgII} phase:  $\log U  >-3.4$.  [{\HI}
cannot address multiphase; {\CIV} requires multiphase.]

\subsection{\rm S18  (Q$0454+039$; $z_{\rm abs}=0.93150$)}

[$N({\FeII})$ measured; $W_{r}({\Lya})$ measured; no break at Lyman limit;
$W_{r}({\CIV})$  upper limit.]   This  system has  measured $N({\FeII})  >
N({\MgII})$,  which a  solar  abundance pattern  and  Haardt \&  Madau
spectrum cannot produce.  However, because only one {\FeII} transition
was observed, the VP fit may  be systematically large due to the noise
characteristics  of the  data (more  so than  the formal  errors would
indicate).  Models will converge  for $N({\FeII})$ at least $1\sigma$
less  than   measured.   We  quote   constraints  for  the   range  of
$N({\FeII})$ reduced by $1\sigma$  to $2\sigma$.  The {\Lya} profile
constrains the metallicity  $-1<Z<0$; the lack of a  Lyman limit break
requires $Z>-1$.  
[As for S7, the constraint differs somewhat from \citet{cwc-lb98}
because of the use of a different version of the FOS/{\it HST} spectrum.]
The $b({\HI})$ from the curve of growth needed to
produce the observed $W_{r}({\Lya})$ (see Figure~\ref{fig:cog})
is consistent with the measured $b({\MgII})$.
These models also  give $-4.7< \log U <-3.6$ for the
full metallicity  range modeled ($-2.5<Z<0$) and $-4.7<  \log U <-3.8$
for $-1<Z<0$.   These models have sizes below  $2.1$~pc.  Larger sizes
($5$ to $175$~pc) would result if iron were enhanced  by $0.5$~dex.  In the
``iron--enhanced''  scenario,  {\Lya}  constrains the  metallicity  as
above,  but  the  Lyman  limit  break is  slightly  less  restrictive,
$Z>-1.2$,  and  the ionization  parameter  is  higher:  $-3.7< \log  U
<-3.3$.  [{\HI} does not require  multiphase; {\CIV} limit is too poor to
address multiphase.]
  
\subsection{\rm S19  (Q$1206+456$; $z_{\rm abs}=0.93428$)}

[$N({\FeII})$  upper limit;  $W_{r}({\Lya})$ measured;  no break  at Lyman
limit; $W_{r}({\CIV})$  measured.]  If all the {\Lya}  absorption arose in
the {\MgII}  phase, this constrains the  metallicity to be  $-2 \leq Z
\leq  -1$.  This  does not  conflict  with the  constraint that  $\log
N({\HI})<16.8$~{\cmsq},  as required  by  the lack  of  a Lyman  limit
break\footnote{The apparent break is entirely accounted for by the
$z=0.9276$ system along this same line of sight \citep{1206}.}.
Thus, a second phase of  gas is not required to explain {\Lya}
and the Lyman limit.  [The value of $b({\HI})\sim20$~{\kms} from the curve
of growth in Figure~\ref{fig:cog} is consistent with
$b({\MgII}) = 7.5$~{\kms} for the model temperature.]
The  {\FeII} limit requires $\log U >-3.7$.  The
{\CIV} detection  can be produced  in the {\MgII}  cloud if $\log  U =
-1.7$;  however, this  slightly underproduces  {\Lya} and  requires an
unreasonably  large  ($80$~kpc)  cloud.  Thus,  multiphase  ionization
structure is required.  [{\HI} does not require multiphase; multiphase
is required to explain {\CIV}.]

\subsection{\rm S20  (Q$0002+051$; $z_{\rm abs}=0.95603$)}

[$W_{r}(2600)$  upper limit;  $W_{r}({\Lya})$ measured;  no break  at Lyman
limit;  $W_{r}({\CIV})$   measured.]   If  all  the   {\Lya}  and  {\MgII}
absorption  arose in  the same  phase, then  the {\Lya}  profile would
require $Z<-2.0$,  which corresponds to  $\log N({\HI})>17.3$~{\cmsq}.
Contradicting  this is  the  absence  of a  Lyman  limit break,  which
requires  $Z \geq -1$.   Thus, a  second phase  with a  larger Doppler
parameter  is needed  to create  the {\Lya}  absorption
(see Figure~\ref{fig:cog}).   The {\FeII}
limit  constrains the  {\MgII} phase  to $\log  U >-3.7$.   The {\CIV}
absorption cannot arise in the  {\MgII} cloud, even for $\log U=-2.0$,
the  highest ionization parameter  for which  Cloudy can  generate the
observed $N({\MgII})$.  Thus, the  {\CIV} also requires a second phase
with large  Doppler parameter, which  is more highly ionized  than the
{\MgII} phase. [Multiphase is required to explain {\HI} and {\CIV}.]

\subsection{\rm S24  (Q$1213-003$; $z_{\rm abs}=1.12770$)}

[$N({\FeII})$ upper limit; no FOS  data.]  No {\it HST\/} spectra were
available for this quasar, so the metallicity of this system cannot be
constrained.   We  doubt  the  veracity of  the  $3.5\sigma$  {\FeII}
detection, since two equally strong lines are  detected within $120$~{\kms}
of the  putative {\FeII} $\lambda  2383$ line.  Were  the {\FeII}
detection  real,  it would  predict  $-5.2<  \log  U <-3.9$  over  the
metallicity  range  $-2.5<Z<0$.  Using  the  {\FeII}  as  a limit,  we
constrain  $\log  U  >  -4.6$.    [No  FOS  data, so multiphase
cannot be addressed.]

\subsection{\rm S25  (Q$0958+551$; $z_{\rm abs}=1.21132$)}

[$N({\FeII})$ upper  limit; $W_{r}({\Lya})$  upper limit; Lyman  limit and
{\CIV} not  covered.]  ~{\Lya} is  blended with $\lambda 1550$  from a
possible {\CIV} doublet at $z=0.7330$ \citep{archiveI}, and the Lyman
limit  is  not covered,  so  metallicity  cannot  be determined.   The
{\FeII}  limit constrains  $\log  U >  -3.5$.   [Multiphase cannot  be
addressed.]

\subsection{\rm S28  (Q$0958+551$; $z_{\rm abs}=1.27238$)}

[$N({\FeII})$ measured; $W_{r}({\Lya})$ measured; Lyman limit not covered;
$W_{r}({\CIV})$ measured.]  If all the  {\Lya} arose in the {\MgII} cloud,
$-2.5  \leq Z  \leq  -2.0$  would be  required.   However, the  {\Lya}
profile  is  unphysically  shaped  and therefore  this  constraint  is
somewhat untrustworthy.   Detected {\FeII} and  {\MgII} require $-3.7<
\log U  <-2.8$, which  cannot arise  in the same  phase as  the strong
detected  {\CIV}.   Thus,  a  second,  more highly  ionized  phase  is
required.    [{\HI}  cannot   address   multiphase;  {\CIV}   requires
multiphase.]


\section{Summary of Cloud Properties}
\label{sec:sumcloud}

In \S\ref{sec:cloudnotes}, we presented constraints on the physical
conditions of fifteen\footnote{There are actually $16$ single--cloud weak
{\MgII} systems  in the  sample.  Unfortunately, no  transitions other
than {\MgII}  were covered for  S1, so photoionization  models for
this system  cannot be constrained.  While we include this  system in
number--of--cloud statistics, we exclude  it from other analyses.}
single--cloud weak {\MgII} systems.  In this section, we summarize 
the inferred properties of weak {\MgII} absorbers.
The inferred upper and lower limits on the metallicities, ionization
parameters, densities, and sizes of the clouds are given in
Table~\ref{tab:tab2}.  As described in point (8) below,
for at least seven of the systems we infer that two phases of gas
are required to simultaneously fit the {\MgII}, {\FeII}, {\CIV},
{\Lya}, and Lyman series absorption.
In Table~\ref{tab:tab2}, and except where  noted below, the inferred
properties are  for the low--ionization {\MgII}  phase; the properties
of  the high--ionization phase  are not  well constrained  because the
relevant transitions were covered only at low resolution.
Figure~\ref{fig:z} presents an overview of the metallicity constraints
on the low ionization phase, while Figure~\ref{fig:dens_ip}
summarizes the constraints on its ionization parameter/density.

The following points characterize the basic measured and inferred
statistical properties of the sample of single--cloud weak {\MgII} absorbers:

1) {\it Single Clouds:} Two--thirds of weak {\MgII} absorbers are
single--cloud systems, as contrasted with strong {\MgII} absorbers,
which are consistent with a Poissonian distribution with a median of seven
clouds per absorber (see Figure~\ref{fig:nclouds}).
This suggests that weak {\MgII} absorbers represent a distinct
population of objects, as does their lack of Lyman breaks.  To produce
predominantly single--cloud systems, weak {\MgII} absorbers should
have a preferred geometry or small covering factor if they arise in
extended galaxy halos or galaxy groups, such that a line of sight is
unlikely to intersect multiple clouds.

2) {\it Doppler Parameters:} Most clouds are unresolved at
$R=6.6$~{\kms}, with Doppler parameters of $2$--$7$~{\kms}.  A few of
the {\MgII} profiles have slightly larger Doppler parameters and
are slightly asymmetric, suggesting blended subcomponents or bulk motions.

3) {\it Metallicities:} As shown by Figure~\ref{fig:z}, the {\MgII}
phases of weak absorbers have metallicities at least one--tenth
solar.  In no case is a cloud metallicity constrained to be lower.  In
fact, in cases where a strong {\Lya} profile would seem to require low
metallicity, assuming a single phase, lack of a Lyman limit break
requires $\log N({\HI}) < 16.8$~{\cmsq} and thus high metallicity.  In
such cases much of the {\Lya} equivalent width arises in a broader,
more highly ionized, second phase of gas.  Thus, the {\MgII} phases
have $\log N({\HI})\sim16$~{\cmsq} and have been substantially
enriched by metals.

4) {\it Ionization Parameters/Densities:} Weak {\MgII} absorbers have
a wide range of ionization parameters, ($-5<\log U<-2$) if a solar
ratio of $\alpha$/Fe is assumed, as shown in Figure~\ref{fig:dens_ip}.
This translates to a range of densities, ($-3.5<\log n_{\rm
H}<0$~{\cmsq}).  There is a degeneracy between $\alpha$--process
enhancement and ionization parameter.

5) {\it Iron--Rich Systems:} $N({\FeII})$ was detected in four of the
$15$ systems.  Detection of iron considerably tightens constraints 
on ionization conditions (and thus sizes and densities), as can be 
seen in Table~\ref{tab:tab2}.  Three systems (S7, S13, and S18) have 
$N({\FeII})\simeq N({\MgII})$.  We term these ``iron--rich'' systems, 
and infer $\log U \sim -4.5$ and $\log n_{\rm H} \sim -1$~{\cc}.
If these clouds were $\alpha$--group enhanced relative
to solar, they could not produce the observed high
$N({\FeII})/N({\MgII})$ ratio, as Figure~\ref{fig:femg-grid}
illustrates.  Therefore, we infer that $[\alpha/{\rm Fe}] < 0$
for these three systems.  Also, since Fe depletes more readily
than Mg \citep{savagearaa,laur}, these systems do not
have significant dust depletion.

6) {\it Systems With {\FeII} Limits:} For most of the clouds without
detected {\FeII}, we infer $\log U$ $>-4$ and $\log n_{\rm H} <
-1.5$~{\cc}.  As Figure~\ref{fig:femg-grid} shows, six are constrained
to have $N({\FeII})$ at least $0.5$ dex below $N({\MgII})$, and thus
are significantly different (either more highly ionized and less dense, or
$\alpha$--group enhanced) than the ``iron--rich'' clouds.  For five of the
clouds (with small $N({\MgII})$) the upper limits on
$N({\FeII})/N({\MgII})$ are not restrictive.

7) {\it Cloud Sizes:} Together, the $N_{\rm H}$ constraint (equivalent
to a metallicity constraint) and the density constraint (equivalent to
an ionization parameter constraint) allow estimation of the cloud
thicknesses, $N_{\rm H}/n_{\rm H}$.  These constraints are given in
Table~\ref{tab:tab2}.  Iron enhancement would increase cloud sizes.
For the three iron--rich clouds (see point 5
above), low $N_{\rm H}$ (high metallicity) and high density (low
ionization) indicate that the clouds are small, $\sim 10$~pc.  For
those systems with only limits on $N({\FeII})$, densities are not
sufficiently constrained to infer sizes.  These clouds may be as small
as the iron--rich clouds, or as large as would be feasible for a
cloud with the observed $b({\MgII})$ of several {\kms} (perhaps
several kpcs).

8) {\it Multiple Phases:} Seven of the $15$ systems require two phases
of gas: the low ionization, narrow {\MgII} phase (which is present by
definition in all systems and which may or may not have detectable
{\FeII}); and a second, ``broader'' phase that gives rise to most of
the {\CIV} and {\Lya} absorption.  This second phase should be rather
highly ionized because broad {\MgII} absorption is not seen.  The
high--ionization phase is required in three systems by {\Lya} and the
Lyman limit, and in five systems by strong {\CIV}.  For S20, it is
required by both {\Lya}/Lyman limit and {\CIV}.

Of the four systems with detected {\FeII}, S7 and S28 have strong
{\CIV} which indicate multiphase conditions; S13 has no {\CIV}
spectral coverage, and S18 has poor {\CIV} limits.  Only in S18 were
{\Lya} and the Lyman limit both covered, and they do not indicate a
second phase.

For systems with $N({\FeII})$ upper limits, three systems (S17, S19,
S20) have {\CIV} absorption that require two phases, three systems
(S12, S15, S16) have {\CIV} $3\sigma$ upper limits more stringent than
the detections, and the remaining six have poor limits or no {\CIV}
coverage.  In S3, S15, and S20, {\Lya} and the Lyman limit indicate
multiphase conditions.  For all these systems, absorption from the
second phase varies considerably in strength, indicating that in some
cases a second phase may be absent, or at least very weak.  The {\Lya}
and {\CIV} profiles are of insufficient resolution to significantly
constrain the properties of the higher ionization phase.

9) {\it Relationships Between Properties:} Column density of {\MgII},
Doppler parameter of {\MgII}, metallicity, ionization parameter, and
presence of a second phase are not found to be correlated properties
in this small sample.


\section{Discussion}
\label{sec:discussion}

\subsection{High metallicity {\Lya} forest clouds}
\label{sec:forest}

Numerical simulations and observations have suggested that the higher
column density, $z<1$ {\Lya} forest clouds arise within a few
hundred kpcs of galaxies \citep{dave,cen,ortiz}.  By
stacking the spectra of $15$ quasars, \citet{barlow}
detected {\CIV} and derived $[{\rm C/H}] \geq
-1.9$ for {\Lya} forest clouds at $z\sim0.5$ (the limit is increased
to $-1.3$ if clustering is considered).  In contrast, at $z=3$, $\log
N({\HI})>15$ {\Lya} forest clouds are seen to have a lower
metallicity, with $[{\rm C/H}]\sim-2.5$ \citep{songaila}.  These
observations indicate that larger column density, sub--Lyman limit
forest clouds were enriched from $z \simeq 2$ to $z \simeq 1$.

As Paper~I\nocite{weak} demonstrated, the relative redshift number
densities of weak {\MgII} and Lyman limit systems require almost all
weak systems to be optically thin in neutral hydrogen with $\log
N({\HI})<16.8$~{\cmsq}.  For five single--cloud weak {\MgII} systems
the FOS spectra cover the Lyman limit; in each of these cases
there is no Lyman limit break \citep{archiveII}, corroborating
the statistical number density argument.  At log $N({\HI})= 16.8$~{\cmsq},
for the measured $N({\MgII})$ of weak systems, the model cloud
metallicities are about $10$\% solar.  Unless the clouds have
super--solar metallicity, $\log N({\HI})$ cannot be much more than a
dex below $16.8$~{\cmsq}.  Therefore, neutral column densities of weak
systems approximately range over $15.8 < \log N({\HI}) < 16.8$~{\cmsq}
(within a few tenths of a dex).

We integrated the column density distribution of {\Lya} forest clouds
over this range of $N({\HI})$ to estimate what fraction
of this portion of the {\Lya} forest is associated with metal enriched
weak systems.  At $z\leq1.5$, only the {\Lya} equivalent width distribution
is measured directly \citep{weymann98}.  Converting equivalent widths
to column densities introduces considerable uncertainty in the slope
$m = d \log N/d \log N({\HI})$.  Using the $m = -1.3$ slope determined
by \citet{weymann98}, we find $dN/dz \sim 4$ for {\Lya}
over the range $15.8 <\log N({\HI}) < 16.8$~{\cmsq}.  Alternatively,
using two power laws, with slopes $m_1 = -1.8$ and $m_2 = -0.6$, which
intersect at $\log N(HI) = 16$~{\cmsq} (also consistent with
\citet{weymann98}), we find $dN/dz \sim 1$ for
the same $N({\HI})$ range.

By comparison, single--cloud weak systems have $dN/dz=1.1\pm0.06$
(computed as in Paper~I).
Thus, for the single power--law and double power--law fits,
respectively, $\sim25$\% and $\sim100$\% of the {\Lya} forest clouds
with $15.8 <\log N({\HI}) < 16.8$~{\cmsq} should be weak {\MgII} systems. 
Based upon our photoionization modeling, this implies that this portion of
the {\Lya} forest has been significantly metal enriched, with $Z \geq
-1$.  In the iron--rich systems, it is likely that the Fe to Mg
ratio is greater than or equal to solar,
which implies enrichment by Type Ia as well as Type II supernovae
\citep{laur,mcwilliam}, which has implications
that we discuss in \S~\ref{sec:highfe} below.

In Paper~I, we showed that $\sim 7$\% of the {\Lya} forest systems with
$W_{r}({\Lya}) \geq 0.1$~{\AA} are weak {\MgII} systems; we have now shown that
weak systems comprise a substantial fraction, {\it perhaps most},
of the $z < 1$ {\Lya} forest with $\log N({\HI}) \sim 16$~{\cmsq}.

\subsection{Space Density of Weak {\MgII} Clouds}
\label{sec:numbers}

Using the sizes derived from Cloudy and the observed redshift number
densities, we can estimate the space density of weak systems in the
universe.  From this we then can deduce the relative numbers of weak
systems compared to galaxies (i.e.\ strong {\MgII} absorbers), regardless
of how they are distributed in the universe.

In general, the space density of absorbers is given by
\begin{equation}
n(z) = \left({{dN}\over{dz}}\right) {{H_0}/c\over{\pi  R_{\ast}^2 C_f
}} \ {{(1+2q_0z)^{1/2}}\over{1+z}},
\label{eq:ndens}
\end{equation}
where $R_{\ast}$ is the characteristic radius of the absorber and
$C_f$ is the covering fraction, such that $\pi R_{\ast}^2 C_f$ is the
effective cross section of an absorber.  We consider the {\MgII} phase only,
since the high--ionization phase is poorly constrained.  Since $n$
is dependent upon the square of the characteristic radius of the absorber,
we consider the iron--rich systems separately because their sizes are
well constrained.\footnote{It is not clear if iron--rich weak systems
are a separate population of absorbers; as Figure~\ref{fig:femg-grid}
shows, we cannot tell whether weak systems are bimodally or
continuously distributed in $N({\FeII})/N({\MgII})$, and thus in
ionization condition and/or $\alpha/{\rm Fe}$ enhancement.  For the
systems with lowest $W_{r}(2796)$, even $N({\FeII})\simeq N({\MgII})$
would place {\FeII} below our detection threshold. Accordingly, the derived
space density of iron--rich systems should be taken as a lower
limit.}  The systems with no measured $N({\FeII})$ have $dN/dz = 0.75 \pm 0.05$
over the same redshift range, and their sizes are relatively
unconstrained.

Using $q_0 = 1/2$, $z=1$, and $C_f=1$, we find $n = 1.3 \times 10^7
h (1{\rm pc}/R_{\ast})^2~{\rm Mpc}^{-3}$ for the iron--rich clouds.
For $R_{\ast} \sim 10$~pc, $n = 1.3 \times 10^5 h~{\rm Mpc}^{-3}$
(the linear dependence on $h$ arises because our sizes are derived
independent of cosmological model).  For the clouds without detected
iron, we find
$n = 56 h~(1{\rm kpc}/R_{\ast})^2~{\rm Mpc}^{-3}$.  A lower limit
can be estimated by assuming that the clouds are not Jeans unstable and
are less than 50~kpc in size.  This yields $n > 0.02 h~{\rm Mpc}^{-3}$.
Recall that $n \propto C_{f}^{-1}$, so a smaller covering factor would
result in a higher space density.

For comparison, strong absorbers have $dN/dz = 0.91 \pm 0.1$
\citep{ss92}, $R_{\ast}\simeq 40h^{-1}$~kpc, and unity $C_{f}$ at
$\left< z \right> = 0.9$,
yielding $n = 0.04h^{3}~{\rm Mpc}^{-3}$ \citep{sdp94}, which is
consistent with the space density of galaxies at this redshift
\citep{lilly95}.  The ratio of the space density of weak {\MgII}
absorbers to the space density of strong {\MgII} absorbers (bright galaxies)
is a simple comparison of the relative numbers of these populations in 
the universe, independent of where they arise or how they cluster.  
Therefore, though we quote the result as a number of weak {\MgII} 
absorbers per $\sim L^{\ast}$ galaxy, this does 
\emph{not} imply that weak {\MgII} absorbers are clustered around or 
associated with bright galaxies.  For the cloud size estimates of 
the previous paragraph, the lower limits on the ratio 
of space densities of weak {\MgII} systems to galaxies are
$\simeq 3 \times 10^{6}h^{-2}$ and $0.5 h^{-2}$ for iron--rich clouds
and clouds without detected iron, respectively.  Again, recall that
$n$ for the weak systems goes as $C_{f}^{-1}R_{\ast}^{-2}$, so a
smaller covering factor and smaller sizes result in higher ratios.

The sample of three iron--rich systems is small, but since space density
depends only linearly on $dN/dz$ (equation \ref{eq:ndens}), for an unbiased
survey this uncertainty does not change the qualitative result; more
important is the robustness of their inferred size (see \S~\ref{sec:robust}).
\emph{This is an astonishing result: iron--enriched single--cloud weak systems
outnumber bright galaxies by at least a million to one.}

\subsection{Relationship to Galaxies and Covering Factor}
\label{sec:galaxies}

As calculated above, the redshift number densities of all
single--cloud weak systems is $\simeq 1$, which is comparable to that
of the strong systems associated with bright galaxies.  Taken at face
value, one could argue that almost all weak systems, though
significantly more numerous than galaxies, are nonetheless associated
with galaxies.  In fact, strong {\MgII} absorbers commonly have weak
clouds with $W_{r}(2796) \simeq 0.1$~{\AA} at intermediate to high
velocities, i.e.\ $40$ to $400$~{\kms} from the absorption systematic
velocity zero point \citep{datapaper}.  We present a direct
comparison of these properties in Figure~\ref{fig:outlying}, which
shows that the ranges of {\MgII} column densities,
Doppler parameters, ionization conditions and $N({\FeII})/N({\MgII})$
are similar.

Thirteen of the single--cloud weak systems in our sample are in QSO
fields that have been imaged in efforts to identify the galaxies
hosting strong {\MgII} absorption.  We cite a few examples
to illustrate that there is growing
evidence against the association between {\it bright\/} galaxies and
{\it single--cloud\/} weak systems (also see Paper~I).

Of the iron--rich systems, two (S7 and S18) are seen in absorption
against Q~$0454+039$, whose field has been well studied
\citep{q0454dla,lebrun97,cwc-lb98}.  No candidate
galaxies are seen at the weak system redshifts within $\sim
10${\arcsec} of the QSO, down to $\sim 0.01 L_B^{\ast}$
\citep{lebrun97}.  The line of sight to Q~$0002+051$ has five {\MgII}
systems, of which three are single--cloud weak systems (Paper~I).  The
strongest (S6), with $W_{r}(2796) = 0.29$~{\AA}, is at the redshift of
a bright galaxy, while the two others, S15 and S20, are unmatched with
galaxies out to $20${\arcsec}.  The line of sight to Q~$1421+331$ has
four {\MgII} systems, of which two are single--cloud weak systems (S1
and S13, the third iron--rich system).  There is one bright galaxy in
the field with an unconfirmed redshift; statistically, it is likely to
be associated with one of the two strong {\MgII} absorbers
\citep{sdp94,steidel95}.

This evidence that weak systems do not commonly select bright galaxies within
$\simeq 50h^{-1}$~kpc of the absorbing gas does not rule out either
dwarf galaxies (or smaller mass objects) or bright galaxies within
$100$--$200$~kpc of the QSO.  That is, they could arise in low luminosity, small
mass structures that are either clustered within a few hundred kpc of
bright galaxies or are distributed in small groups of galaxies
(analagous to the Local Group).  If so, this might lead us to consider
whether the intermediate and high velocity weak {\MgII} clouds in strong systems
do not reside within $\simeq 50h^{-1}$~kpc of galaxies, but instead are the
same objects as the weak systems; that is, the weak clouds in strong
systems sampled by the lines of sight that select bright, {\MgII}
absorbing galaxies could in principle be interloping weak systems that
arise throughout the group environment.  However, as we will show,
there are problems with this scenario.

If the single--cloud weak {\MgII} absorbers and the kinematic outliers
of strong systems are the same population (with the same
spatial distribution and covering factor) then they should have
similar redshift number densities.  This should be true regardless
of whether they are clustered within $100$--$200$~kpc of galaxies
or distributed throughout groups.
If they are not the same population of objects, then
their  relative  redshift number densities  gives  the  ratio of  their
absorption  cross  sections,  
\begin{equation}  
\frac{(dN/dz)_{w}}{(dN/dz)_{o}}   =  \frac{\sigma_{w}}{\sigma_{o}}   
=   \frac{(C_{f}  n R_{\ast}^{2})_{w}}{(C_{f} n
R_{\ast}^{2})_{o}} ,
\label{eq:cross_sec}
\end{equation}
where the terms are the same as described in Equation~\ref{eq:ndens},
and where $w$ and $o$ denote weak systems and kinematic outliers of
strong systems, respectively.

In strong {\MgII} absorption systems, the chance of intercepting one 
or more intermediate or high velocity, single--weak outliers is 
roughly $55$\%; in
roughly half of those systems, two or three single--weak clouds are
observed \citep{datapaper}.  This translates to a redshift number
density of $(dN/dz)_{o} \sim 0.7$, regardless of the spatial
relationship between galaxies and the intermediate and high velocity
outliers.

$L^{\ast}$ galaxies cover only a very small fraction of a group;
consequently, passing through two strong systems or a strong system
and a weak system should be similarly improbable, unless weak
absorbers cluster strongly around bright galaxies.  If the weak
{\MgII} absorbers were clustered within $100$--$200$~kpc of strong
{\MgII} absorbers, a candidate $L^*$ galaxy would be observed, within
$40 h^{-1}$~kpc, $5$--$10$\% of the time, which does not violate the
observational constraint.
However, in this scenario, since the weak absorbers are spread over
a fairly large region, their covering factor within that $100$--$200$~kpc
radius is also fairly small, and so one would only expect a high
velocity weak outlier in $10$--$20$\% of strong systems.  This conflicts
with the observed fraction, which is close to $100$\% 
(some systems have multiple outliers.)

Thus, to explain the kinematic outliers of strong systems, the weak
{\MgII} absorbers would need to be strongly clustered around the 
strong absorbers.  They would be rare at large distances.  However,
they would also be rare at small distances, or
else more candidate galaxies for weak {\MgII} 
absorbers should have been found within $\simeq 50 h^{-1}$~kpc.
This clustering, only at intermediate distances, seems rather contrived.  
A further complication is that clouds within $100$~kpc
should merge with the nearby galaxy on timescales of $<10^9$~years.
 
We therefore conclude that while weak {\MgII} absorbers and outliers
of strong {\MgII} absorbers are similar in their physical properties,
they generally arise in different types of hosts.  In the case of the
outliers, the host is apparently the bright galaxy responsible for the
strong {\MgII} absorption.  The high velocity clouds around this
galaxy could arise from gas tidally stripped from companions, or
through energetic events which eject gas from the disk
\citep{archivelet}.  The processes which give rise to the weak
{\MgII} absorbers may be similar, but may occur in less massive, less
luminous hosts.

\subsection{Possible Environments of Weak {\MgII} Absorption}

If we take the inferences presented above as reasonable approximations
for the true physical conditions of weak {\MgII} absorbers, then what are
the implications?  In what environments does weak {\MgII} absorption
arise?

\subsubsection{Iron--rich clouds}
\label{sec:highfe}

We now consider the environments of the iron--rich weak {\MgII}
absorbers (with $dN/dz \sim 0.2$).  These absorbers, with comparable column
densities of {\FeII} and {\MgII}, are inferred to be high metallicity 
($> 0.1~Z_{\odot}$), small [$\log N({\HI})<16.8$~{\cmsq} and size
$\sim10$~pc] gas clouds, which, if spherical, would outnumber bright
galaxies by a factor of $\sim10^6$.
They seem not to be closely associated with bright galaxies, i.e.
within $\simeq 50 h^{-1}$~kpc of $\ga 0.05~L^*$ galaxies.

The small inferred gas masses (a few $M_{\odot}$) and small velocity
dispersions ($b\sim6$~{\kms}) of the {\MgII} phase in the iron--rich
systems suggest objects that would not be stable over astronomical
timescales.  Either the gas is transient or the clouds are confined
by outside gas pressure or stellar and/or dark matter.  The
{\MgII} phase would be a condensation inside a larger structure,
which may give rise to the higher ionization, larger Doppler
parameter phase.  However, it is important to note that these two phases
would not be in simple pressure equilibrium, as they have similar 
inferred temperatures (if both are photoionized) and different inferred
densities.

The high metallicities ($\geq0.1$ solar) of weak {\MgII} absorbers
require substantial enrichment.  Yet because the gas in the iron--rich
systems is not $\alpha$--group enhanced relative to solar, it cannot have
been enriched solely by Type II SNe or galactic winds produced by
multiple Type II SNe, as material processed in this way is observed to
be $\alpha$--group enhanced by 0.5 dex \citep{mcwilliam,laur,tsuji}.
Therefore, weak {\MgII} absorbers with high {\FeII} must have retained 
enriched gas from Type Ia SNe.  In general, retaining high--velocity 
SNe ejecta requires either a deep
potential well or ``smothering'' of the explosion by gas in the
surrounding medium.  For weak {\MgII} absorbers, the lack of Lyman
limit breaks and associated bright galaxies within $\simeq 50 h^{-1}$~kpc argues
that $L^{\ast}$ galaxy potential wells are not responsible.
Accordingly, we consider how Type Ia supernova ejecta might be
retained by smaller potential wells.

In the pre--dark matter era, \citet{peebles} explored
the formation and expected properties of Population III star clusters.
We explore the updated general scenario of a dark matter mini--halo of
$10^6$--$10^8$~M$_{\odot}$ with a virial velocity of tens of {\kms}.
Such a mini--halo could contain a dwarf galaxy or only a star cluster
within it \citep{rees}.  When the first massive stars in the cluster exploded as 
Type II SNe, the resulting superbubble would have driven much of the 
surrounding gas out into the halo.  This process should destroy small 
halos, which sets a lower limit on the mass of halos which survive.  In
sufficiently large halos, the superbubble gas should be slowed as it
sweeps out into the halo gas.  Eventually the shell should slow to the
virial speed of the halo, and may cool and fragment or mix with the
halo gas and disperse.  As the product of Type II SNe, this gas should
be $\alpha$--group enhanced.  Condensations within the superbubble
remnant might give rise to detectable {\MgII} but not {\FeII}
absorption.

After the requisite delay time ($\sim1$~Gyr), Type Ia SNe should
detonate within the star cluster.  Scaling roughly,
$10^5$~M$_{\odot}$ in stars should produce one Type Ia supernova per 
$10^{9}$~yrs, assuming that the Milky Way SNe rate scales to
lower mass structures.
If this Type Ia supernova gas were retained, mixed with already
$\alpha$--group enhanced gas, and condensed, such a structure might be
observed as an iron--rich, high--metallicity weak {\MgII} absorber.
In the absence of a large potential well, trapping the debris would
require smothering, either at small radii within a parent star
cluster, or at larger radii within the surrounding halo.  If the Type
Ia ejecta were trapped within a star cluster, the observed small
{\MgII} Doppler parameter would represent the low virial speed of the
cluster; if the ejecta were trapped within the halo, the small
$b({\MgII})$ would indicate a condensation in the supernova shell. 
\citet{burkert} have also considered 
the effect of Type Ia SNe on the gas in dwarf spheroidals.

To summarize, the high inferred metallicities and lack of nearby bright
galaxies imply that weak {\MgII} absorption arises in metal--enriched gas inside
small dark matter halos.  To consider what type of luminous structures could 
exist inside the halos (dwarf galaxies or star clusters)
and how the absorbing gas is distributed within the halo (concentrated within 
star clusters or at large in the halo), we must balance
the two factors that determine the absorption cross section:
number of parent halos per $L^*$ galaxy and the number of absorbers within
these halos.  Physically, the latter factor is determined by the generation
rate of the absorbing gas and the persistence of the structure.

Equation \ref{eq:cross_sec} simply relates absorber sizes and number densities
of two populations to the ratio of their redshift number densities.  Because
the absorption statistics of strong {\MgII} absorbers are well--established at
redshift $0.3\leq z \leq 2.2$ \citep{ss92}, and because $L > 0.1~L^*$ galaxies
are largely responsible for this strong absorption, it is useful to compare
them to weak {\MgII} absorbers.  We can rearrange equation \ref{eq:cross_sec}
and use $R_s = 40 h^{-1}$~kpc, $dN/dz_s = 0.91 \pm 0.1$ for strong systems,
$dN/dz_w = 0.18 \pm 0.01$ for iron--rich weak systems, and unity covering
factor $C_f$.  Then, $n_w/n_s$ is the ratio of the number of halos containing
iron--rich weak absorbers to the number of strong absorbers.
The result is that, in each weak absorber halo, weak {\MgII} absorption
covers the same area as a circle with radius $R_w = 17 h^{-1} (n_w/n_s)^{-1/2}$~kpc.

For the Milky Way, if only the dozen known dwarf satellites contribute to
iron--rich weak {\MgII} absorption, then $R_w = 7$~kpc for $h=0.7$.  
Obviously, a very
large fraction of each dwarf must give rise to the absorption in this
scenario.  Simulations generically predict more dark matter halos per poor
group than are observed as dwarf galaxies in the Local Group \citep{klypin,moore}.
For a typical $L^*$ galaxy, simulations by \citet{klypin} of poor
groups produce about 500 dark matter halos with $v_{circ} > 10$~{\kms}
per $L^*$ galaxy.  Using this for $\eta$ yields 
$R_w\sim1$~kpc per small halo for $h=0.7$.  Even with this large population 
of satellites, a large fraction of each halo would need to give rise 
to weak {\MgII} absorption with high $N({\FeII})/N({\MgII})$.

Dwarf galaxies and faint dark matter mini--halos might be expected to cluster
less strongly than brighter galaxies.  If they exist in abundance in voids,
then this would raise the number of weak absorber halos per $L^*$ galaxy, 
$n_w/n_s$, and decrease the effective absorption radius per halo, $R_w$.

If Population III star clusters exist inside numerous small dark
matter halos, gas trapped within the clusters might give rise to weak
{\MgII} absorption. The correspondence between the virial velocity of
a globular cluster and the small Doppler parameter of weak {\MgII}
absorbers is suggestive, as is the similarity between globular cluster
radii and the inferred sizes of the iron--rich weak {\MgII} clouds.
Two problems of this scenario are that sufficiently small halos should
be destroyed by the initial burst of Type II SNe, and that rogue star
clusters with concentrations similar to that of Milky Way globulars
would have been detected in the Local Group.  More diffuse clusters
might remain below present detection thresholds.  The expected
concentration is unknown, as it is difficult to calculate the expected
packaging of the first and second generation of stars to form in the
Universe \citep{rees,abel,abel2000}.

Still, unless there are more than a million mini--halo hosts of
iron--rich weak {\MgII} absorption for every $L^*$ galaxy in the
Universe, the absorption cross section per small halo spans more area
than the inferred $10$~pc size of an iron-rich weak {\MgII} absorber.
Assuming spherical geometries, this would require multiple sites which
could give rise to weak {\MgII} absorption per small halo.  In the
picture where enriched gas is trapped within star clusters, this would
require multiple star clusters per halo.

Alternatively, the gas could exist not in small, isolated structures
but rather in sheets within the halos, which would explain the small
sizes inferred for the absorbers and the large cross section for
absorption per halo.  This would correspond to the general picture
discussed where enriched gas from supernovae, which has fragmented and
cooled, is trapped within a halo.

\subsubsection{Clouds Without Detected {\FeII}}
\label{sec:lowfe}

The sizes of the clouds with upper limits on $N({\FeII})$ are not
constrained.  They could be as small as the iron--rich clouds or much
larger, as large as narrow, single--cloud kinematics permit.  Like the
iron--rich clouds, these clouds with smaller ratios of $N({\FeII})$ to
$N({\MgII})$ arise in $Z \geq -1$ environments.  However, the clouds
without detected {\FeII} \emph{do not} require $[\alpha/{\rm Fe}]\sim 0$;
they could be $\alpha$--group enhanced.  So unlike the
iron--rich clouds, clouds without detected {\FeII} could be wholly
externally enriched.  Given that they are apparently not closely
associated with bright galaxies, two possible origins for their high
metallicities are apparent: external enrichment from larger
structures, or trapping of local SNe ejecta.

In the external enrichment scenario, the winds and superbubbles of
large galaxies pollute the intragroup gas and the low mass structures
in the group: the low mass galaxies, tidal debris, and infalling
clouds.  X-ray observations of poor groups indicate that such gas
would have relatively high metallicities ($>0.1$ solar)
\citep{mulchaey}.  The level of $\alpha$--enhancement depends on the
ability of the galaxy group to retain the early Type II supernova ejecta
which may escape into the intergalactic medium \citep{davis,fino}.
In this case, systems with lower {\FeII} would arise in
galaxy groups, and might be thought of as low neutral column density
(sub--Lyman limit) high velocity clouds (HVCs).  These would not be
analogous to HVCs observed locally in $21$~cm emission, which have
$\log N({\HI})>18$~{\cmsq} and may or may not be of extragalactic origin
\citep{blitz,charlton}.  Rather, they would be more like the
sub--Lyman limit HVCs observed in {\CIV} absorption around the Milky
Way, which are likely extragalactic \citep{sembach}.  Such {\CIV}
HVCs are consistent with a highly ionized single phase, but their
{\SiII} and {\CII} detections are also consistent with what would be
expected for weak {\MgII} absorbers.

Because of the degeneracy between high ionization and
$\alpha$--enhancement, we do not know whether the clouds without
detected {\FeII} are $\alpha$--group enhanced.  If they are not, and instead have solar
$\alpha/{\rm Fe}$, then like the iron--rich clouds, they cannot have
been enriched by $\alpha$--group enhanced external gas.  In this case, the
argument for the high {\FeII} clouds applies: in the absence of a
large potential well, Type Ia SNe ejecta need to be smothered and
trapped by nearby gas.  In this scenario, clouds with lower {\FeII} would be
more highly ionized or $\alpha$--group enhanced versions of the iron--rich
clouds.

If the host star cluster is coeval, like a globular cluster, then
after the initial burst of Type II SNe, all successive SNe should be
Type Ia.  If the stars arise in something more like a dwarf galaxy
with more continuous or stochastic star formation, then Type II SNe
remnants as well as high--ionization pockets of Type Ia SNR could give
rise to lower {\FeII} systems.  Differences in the velocity spread,
state of ionization, and absorption strength of the second phase
seen in many of the weak {\MgII} absorbers might reflect differing
host environments.


\section{Evolutionary Histories and Further Investigations}
\label{sec:further}

\subsection{Weak {\MgII} systems at other redshifts} 

How would these $z\sim1$ weak {\MgII} absorbers appear at higher
redshift?  At $z>1$, the metagalactic ionizing flux was
stronger, rates of star formation and galaxy interaction were higher,
less material had condensed into galaxies, and less time had
transpired for production of Type Ia SNe than at $z=1$.  Thus, clouds 
with the same
total hydrogen column density would have been more highly ionized,
$\alpha$--group enhanced, and lower in metallicity.  (We caution that since the
star formation history of these systems is unknown, it is difficult
to predict metallicity and enhancement evolution.)  Higher ionization
and lower metallicity would make {\MgII} absorption weaker, perhaps
below detection thresholds.  Consequently, sub--Lyman limit weak
{\MgII} absorbers may be rarer or non-existent at high redshift.  
Due to increased
ionization and possible $\alpha$--group enhancement, {\FeII} detections for
weak {\MgII} absorbers should become rarer at $z>1$.

If this population of objects would not be common at high redshift,
what kinds of objects \emph{would} be selected by weak {\MgII}
absorption?  Some weak {\MgII} absorbers at high redshift might still
be below the Lyman limit, but the threshold for detecting {\MgII}
should be pushed to higher $N_{\rm H}$ by metallicity and ionization
effects.  Because of the column density distribution function, this
would lead to fewer absorbers per unit redshift.  However, both
cosmological evolution and the fact that small clouds had not yet
merged into large structures would have the opposite effect.
Regardless of the relative numbers, it is likely that at a
sufficiently high redshift, Lyman limit systems would be detected as
weak {\MgII} absorbers.  Thus, objects physically associated with
weak {\MgII}
absorbers at high redshift may be completely different from those at
$z\sim1$, and may be associated with bright galaxies.

How would the weak {\MgII} absorbers at $z\sim1$ have evolved to the
modern epoch?  Under the less intense metagalactic flux of $z=0$,
detectable {\MgII} absorption can arise in clouds with less neutral
and total hydrogen than at $z\sim1$.  This would make weak {\MgII}
absorbers more common at the present day, since they would extend
further down into the {\Lya} forest.  However, cosmological expansion
and destruction through mergers should have the opposite effect.  Such
low ionization clouds should also have higher {\FeII} then at $z\sim1$, 
since there should have been more time for Type Ia SNe to occur.  This may
not be true for absorbers with recently--formed dwarf galaxy hosts.

\subsection{Future Investigations}

Additional studies at $z\sim1$ can test some of the inferences of this
paper and further constrain the properties of weak {\MgII} absorbers.
Here we briefly discuss four promising avenues: spectroscopy of
{\MgII} absorbers in multiply lensed QSOs, STIS UV spectroscopy,
searches for {\CIV} without {\MgII}, and narrow--band imaging.

1) The critical inference that high {\FeII} clouds have sizes
$\sim10$~pc can be tested by finding weak {\MgII} systems in the
spectra of multiply--lensed quasars.  The best constraint on absorber
sizes thus far was derived from a $z=3.6$ absorption system in
Q$1422+231$, which, due to lensing, is probed by two lines of sight
separated by $13h^{-1}$~pc \citep{rauch}.  This sub--Lyman limit
system has complex absorption spread over 400~{\kms}, observed in both
low and high ionization transitions.  In the low ionization
transitions ({\CII} $\lambda 1334$ and {\SiII} $\lambda 1260$), column
densities vary by a factor of up to 10 between the two lines of sight.
In particular, the reddest component was detected in only one of the
two sightlines.  The inferred density, gas mass, and metallicity of
this component are consistent with the inferred values for high
{\FeII} weak {\MgII} clouds.  Based on Cloudy models, this component
would have detectable weak {\MgII} absorption and $N({\FeII}) \sim
N({\MgII})$.

Size constraints have also been determined directly for {\MgII}
absorbers, though at larger spatial scales.  Eight weak {\MgII}
absorbers have been observed in the $z=3.911$ QSO APM$08279+5255$, in
a very high signal-to-noise spectrum that combined light from multiple
images.  (The two brightest images are separated by $0.35${\arcsec}.)
For 3 systems at $z=1.21$, $1.81$, and $2.04$, {\MgII} $\lambda2796$
and $\lambda2803$ cannot be fit simultaneously with Voigt profiles,
which implies that, due to partial covering of the images, the column
densities are significantly different along the two major lines of
sight, with separations ranging from $0.5$--$1.5 h^{-1}$~kpc
\citep{ellison}.

2) With low resolution UV spectra, the properties of the higher
ionization phase cannot be well constrained.  High resolution spectra
with STIS/{\it HST\/} will soon be available for some of the quasars in this
sample.  This new data should reveal whether the 
{\CIV} resolves into multiple components at $R=30,000$ ($10$~{\kms}),
and whether
the high ionization phase is offset in velocity from the low
ionization phase.  If the low ionization phase is due to ejecta in a
larger halo, we might often see a velocity offset of tens of {\kms}.
For the systems for which {\CIV} was not detected with FOS, we can
determine whether the high ionization phase is truly absent.  With
additional transitions, ionization conditions in the higher ionization
phase can also be determined.

3) Sensitive searches for {\CIV} at $z\sim1$, especially {\CIV} with
no corresponding {\MgII} absorption, would constrain the relative
sizes of the {\CIV} and {\MgII} phases, given the picture that narrow
weak {\MgII} absorption arises in a condensation surrounded by a
broader higher ionization phase.

4) It would be very time--consuming to search for galaxy groups at the
redshifts of known {\MgII} absorbers via wide--field imaging and
spectroscopy.  Narrow--band imaging is more feasible \citep{yanny},
although small, low luminosity galaxies directly in front of the QSO
would still be missed.

Low--redshift investigations may also be relevant.  Weak {\MgII}
absorbers at low redshift should be detected serendipitously in
STIS/{\it HST\/} QSO spectra.
Because of the small redshift path--length for detection of {\MgII}
absorption, few detections are expected.
Nevertheless, such detections would help to constrain the evolution of
$dN/dz$, and at low redshift, searching for associated luminous
structures may be more feasible.  Also, deep 21 cm mapping and absorption
studies should shed light on how high velocity and other Lyman limit
clouds are distributed around nearby galaxies and within nearby
groups, further probing the nature of faint, low--column density 
structures in the universe.

\section{Conclusion}
The basic properties of the single--cloud weak {\MgII} absorbers were
outlined in \S~\ref{sec:sumcloud}.  We conclude the paper by summarizing
our discussion of the nature of these absorbers and their relationship
to other classes of absorbers and objects.

1) Single--cloud weak {\MgII} absorbers are of high metallicity
($Z \ge -1$) and they comprise a large fraction of the
$\log N({\HI}) \sim 16$~{\cmsq} {\Lya} forest (see \S~\ref{sec:forest}).

2) The physical properties of the single--cloud weak {\MgII} absorbers
are similar to those of kinematic outlier clouds in strong {\MgII}
systems.  However, most weak absorbers are not observed within $\simeq
50 h^{-1}$~kpc of $L^{\ast}$ galaxies. Cross--section arguments,
outlined in \S~\ref{sec:galaxies}, indicate that the single--cloud
weak {\MgII} absorbers and the kinematic outlier clouds in strong
{\MgII} systems {\it cannot} be one and the same population of objects
(viewed from different orientations).  They can, however, have a
related process of origin.

3) Three single--cloud weak {\MgII} absorbers are constrained by their
relatively large {\FeII} column densities to have small physical sizes, 
$<10$~pc.  Their observed $dN/dz$, compared to that for strong {\MgII} 
absorbers, indicates that, if that small, they should outnumber 
$L^{\ast}$ galaxies by more than a
factor of a million (see \S~\ref{sec:numbers}).  These iron--rich
single--cloud weak {\MgII} absorbers do not correspond to any known
population of object in the local universe.  As we discuss in
\S~\ref{sec:highfe}, their Fe to Mg ratio
requires \emph{in--situ} enrichment by Type Ia supernovae.  Their sizes
and velocity dispersions suggest an origin in star clusters (the
elusive Population III?) or in shell fragments from the supernovae.
The number of iron--rich {\MgII} absorbers required is large even compared to
the number of low--mass dark matter halos (``failed galaxies'')
predicted by dark matter simulations.

4) The physical properties (particularly the sizes) are not as
well--constrained for the larger subset of single--cloud weak {\MgII}
absorbers without detected {\FeII} (see \S~\ref{sec:lowfe}).  Unlike
the iron--rich population these could be $\alpha$--group enhanced,
though their lack of association with bright galaxies require
energetic ejection or an origin in dwarfs.  The low--iron subclass
could represent lines of sight through sub--Lyman limit regions of
high velocity clouds in galaxy groups.  Alternatively, these low--iron
single--cloud weak {\MgII} absorbers could arise in fragments of Type
II supernovae or in relatively high--ionization fragments from Type
Ia supernovae.

The precise nature of the objects that host single--cloud weak {\MgII}
absorbers is not known.  Generally, they select high metallicity
pockets of material in intra--group and/or intergalactic space. The
phase structure apparent in many of them suggests condensations within
larger potential wells, such as dwarfs, but the large number of
absorbers is surprising.  Understanding the processes of origin of these
mysterious weak {\MgII} absorbers is likely to teach us about a
common, but heretofore unknown, metal--enriched class of object.

\acknowledgements  

Support for this work was provided by the NSF (AST-9617185) and by NASA
(NAG5-6399).  JRR was supported by an NSF REU supplement.  We thank Gary
Ferland for making Cloudy available to the astronomical community.  We are
grateful to more colleagues than we can acknowledge here for stimulating 
discussions during the course of this work, and we give special thanks to 
Alan Dressler, Mike Fall, Jim Peebles, Blair Savage, Ken Sembach, 
Steinn Sigurdsson, and Todd Tripp.



\newpage

\begingroup
\begin{table}
\scriptsize
\tablenum{1}
\label{tab:tab1}
\begin{center}
\begin{tabular}{lcccccccc}
\multicolumn{9}{c}{\sc Table 1: Single Cloud Weak {\MgII} Absorbers}\\
\hline\hline
ID & $z_{abs}$ & QSO & $N({\MgII})$ & $b({\MgII})$ & $N({\FeII})$ & $b({\FeII})$ & $W_r({\CIV})$ & $W_r({\Lya})$ \\
 & & & [{\cmsq}] & [{\kms}] & [{\cmsq}] & [{\kms}] & [{\AA}] & [{\AA}] \\
\hline
S1 & $0.4564$ & $1421+331$ & $13.07\pm0.06$ & $ 7.65\pm0.61$ & \nodata & \nodata & \nodata & \nodata \\
S3 & $0.5215$ & $1354+193$ & $11.91\pm0.05$ & $ 4.88\pm0.90$ & $<11.98$ & \nodata & $<0.24$ & $1.08\pm0.08$ \\
S6 & $0.5915$ & $0002+051$ & $12.63\pm0.01$ & $ 6.78\pm0.22$ & $<11.97$ & \nodata & $<0.23^{a}$ & \nodata \\
S7 & $0.6428$ & $0454+036$ & $12.74\pm0.02$ & $ 5.79\pm0.25$ & $12.60\pm0.05$ & $ 5.31\pm0.88$ & $ 0.38\pm0.03$ & $ 0.70\pm0.05$ \\
S8 & $0.7055$ & $0823-223$ & $12.40\pm0.02$ & $13.30\pm0.62$ & $<11.78$ & \nodata & $< 0.18$ & \nodata \\
S12 & $0.8182$ & $1634+706$ & $12.04\pm0.03$ & $ 2.06\pm0.41$ & $<11.84$ & \nodata &  $< 0.07$ & \nodata \\
S13 & $0.8433$ & $1421+331$ & $13.10\pm0.10$ & $ 3.15\pm0.23$ & $13.47\pm0.07$ & $ 2.34\pm0.18$ & \nodata & \nodata \\
S15 & $0.8665$ & $0002+051$ & $11.89\pm0.04$ & $ 2.65\pm0.82$ & $<11.94$ & \nodata & $< 0.11$$^{a}$ & $ 0.81\pm0.10$ \\
S16 & $0.8955$ & $1241+174$ & $11.73\pm0.06$ & $ 7.51\pm1.44$ & $<11.58$ & \nodata & $< 0.10$ & $ 0.45\pm0.05$ \\
S17 & $0.9056$ & $1634+706$ & $12.47\pm0.01$ & $ 2.77\pm0.10$ & $<11.60$ & \nodata & $ 0.18\pm0.02$ & $ 0.49\pm0.03$ \\
S18 & $0.9315$ & $0454+036$ & $12.29\pm0.08$ & $ 1.52\pm0.19$ & $12.24\pm0.08$ & $ 2.28\pm1.46$ & $< 0.62$ & $0.31\pm0.07$ \\
S19 & $0.9343$ & $1206+456$ & $12.05\pm0.02$ & $ 7.52\pm0.52$ & $<11.48$ & \nodata & $ 0.25\pm0.05$ & $ 0.47\pm0.07$ \\
S20 & $0.9560$ & $0002+051$ & $12.15\pm0.02$ & $ 7.54\pm0.58$ & $<11.58$ & \nodata & $ 0.52\pm0.04$ & $ 0.85\pm0.07$ \\
S24 & $1.1278$ & $1213-003$ & $12.11\pm0.05$ & $ 1.94\pm0.44$ & $<11.96$ & \nodata & \nodata & \nodata \\
S25 & $1.2113$ & $0958+551$ & $12.41\pm0.03$ & $ 3.34\pm0.34$ & $<11.67$ & \nodata &  \nodata & $< 0.92$ \\
S28 & $1.2724$ & $0958+551$ & $12.57\pm0.02$ & $ 3.92\pm0.21$ & $11.99\pm0.22$ & $1.19\pm1.30$ & $ 0.44\pm0.03$ & $0.75\pm0.15$ \\
\hline
\multicolumn{9}{l}{$^{a}$ {|CIV} $\lambda 1550$ equivalent width}
\end{tabular}
\end{center}
\end{table}


\newpage

\begingroup
\begin{table}
\scriptsize
\tablenum{2}
\label{tab:tab2}
\begin{center}
\begin{tabular}{lcccccrrr}
\multicolumn{9}{c}{\sc Table 2: Inferred Properties of Single Cloud Weak {\MgII} Absorbers}\\
\hline\hline
ID & $\log Z>$ & $\log Z<$ & $\log U>$ & $\log U<$ & $\log n>$ & $\log n<$ & $d>$ & $d<$ \\
 & [$Z_{\odot}$] & [$Z_{\odot}$] & & & [{\cc}] & [{\cc}] & [pc] & [pc] \\
\hline\hline
\multicolumn{9}{c}{Clouds With Detected {\FeII}} \\
\hline
S7 & $-1.0$   & $0.0$   & $-4.5$  & $-4.2$  & $-1.4$    & $-1.1$  & $2$      & $8$      \\
S13 & \nodata & \nodata & $-5.0$  & $-3.0$  & $-2.5$    & $-0.8$  & $1$      & $12$  \\
S18 & $-1.0$  & $0.0$   & $-4.7$  & $-3.8$  & $-1.4$    & $-0.5$  & ---      & $2$  \\
S28 & ---     & ---     & $-3.7$  & $-2.8$  & $-2.4$    & $-1.5$  & $10$     & $16000$  \\
\hline
\multicolumn{9}{c}{Clouds Without Detected {\FeII}} \\
\hline
S1 & \nodata  & \nodata & \nodata & \nodata & \nodata   & \nodata & \nodata  & \nodata \\  
S3 & $-1.5$     & ---     & ---     & ---     & \nodata   & \nodata & \nodata  & \nodata \\  
S6 & \nodata  & \nodata & $-3.5$  & ---     & ---       & $-2.1$  & $17$     &  $30000$ \\ 
S8 & \nodata  & \nodata & $-3.6$  & $-2.4$  & $-3.2$    & $-2.0$  & $7$      & $50000$  \\
S12 & \nodata & \nodata & $-4.4$  & ---     & ---       & $-0.7$  & ---      & $3\times10^9$  \\
S15 & $-1.0$  & ---     & $-3.6$  & ---     & ---       & $-1.8$  & $3$      & $3000$  \\
S16 & $-2.5$  & $-1.0$  & $-4.8$  & $-2.0$  & $-3.2$    & $-0.4$  & $0.3$    & $14000$  \\
S17 & $-2.0$  & ---     & $-3.4$  & ---     & ---       & $-1.8$  & $1$      & $50000$  \\
S19 & $-1.0$  & ---     & $-3.7$  & $-1.7$  & $-3.5$    & $-1.5$  & $10$     & $10000$  \\
S20 & $-1.0$  & ---     & $-3.7$  & ---     & ---       & $-1.6$  & $1$      & ---  \\
S24 & \nodata & \nodata & $-4.6$  & \nodata & ---       & $-0.6$  & ---      & ---  \\
S25 & blend   & blend   & $-3.5$  & \nodata & ---       & $-1.7$  & ---      & ---  \\
\hline
\end{tabular}
\end{center}
\end{table}


\newpage
\begin{figure*}
\figurenum{1a}
\plotone{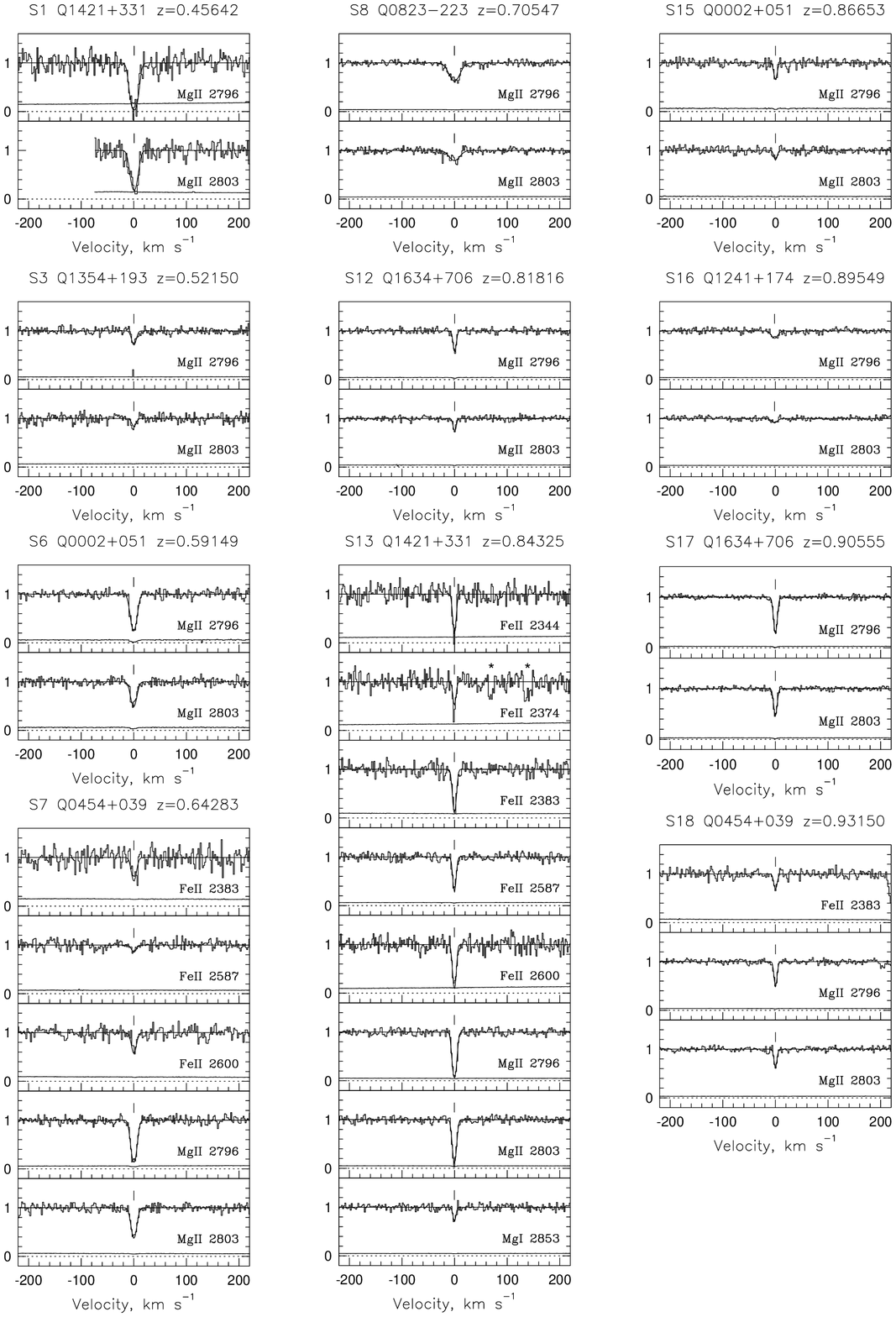}
\end{figure*}

\clearpage

\begin{figure*}
\figurenum{1}
\plotone{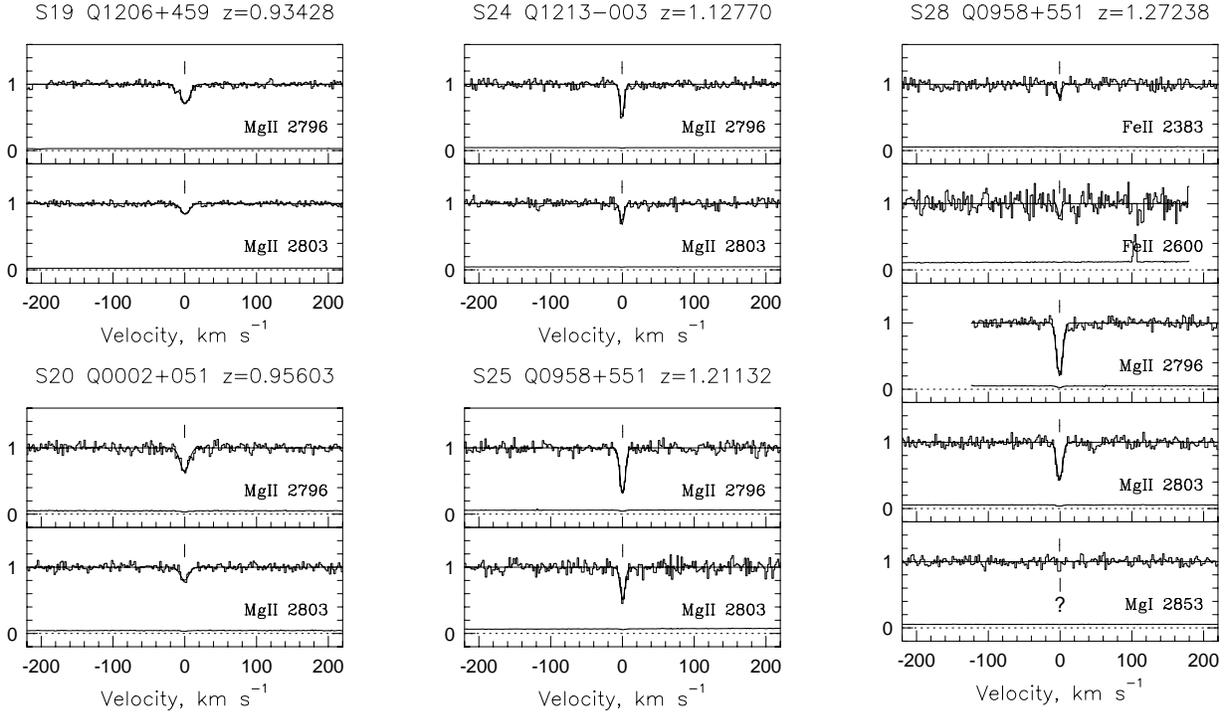}
\protect\caption{
The high--resolution spectra for the 16 single--cloud weak {\MgII}
absorbers of this sample.  {\MgII}, {\MgI}, and {\FeII} were captured 
with HIRES/Keck
at $R=$6.6~{\kms} resolution; MINFIT Voigt profiles are superimposed
(see \S~\ref{sec:spec_vp}).
Table~\ref{tab:tab1} lists the column densities and Doppler
parameters.}
\label{fig:spectra}
\end{figure*}

\clearpage

\begin{figure*}
\figurenum{2}
\plotone{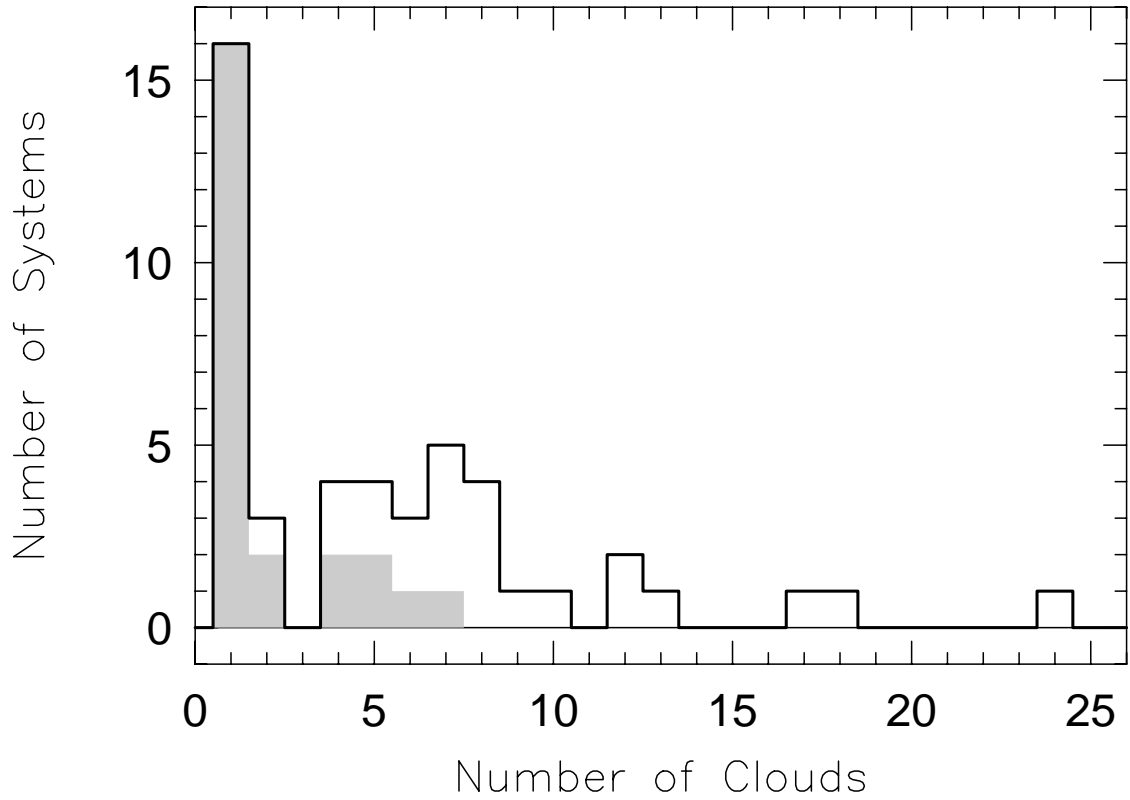}  
\protect\caption{
The distribution of the number clouds per {\MgII} absorption system,
for a limiting equivalent width of $W_{r}(2796) = 0.02$~{\AA}.  The total
distribution (strong and weak combined) is outlined, and weak clouds
are shaded.  Weak absorbers show a strong spike at $N=1$ cloud per
system, which suggests that they are a different type of object than
the strong absorbers.
}
\label{fig:nclouds}
\end{figure*}

\clearpage

\begin{figure*}
\figurenum{3}
\vglue -0.9in
\plotone{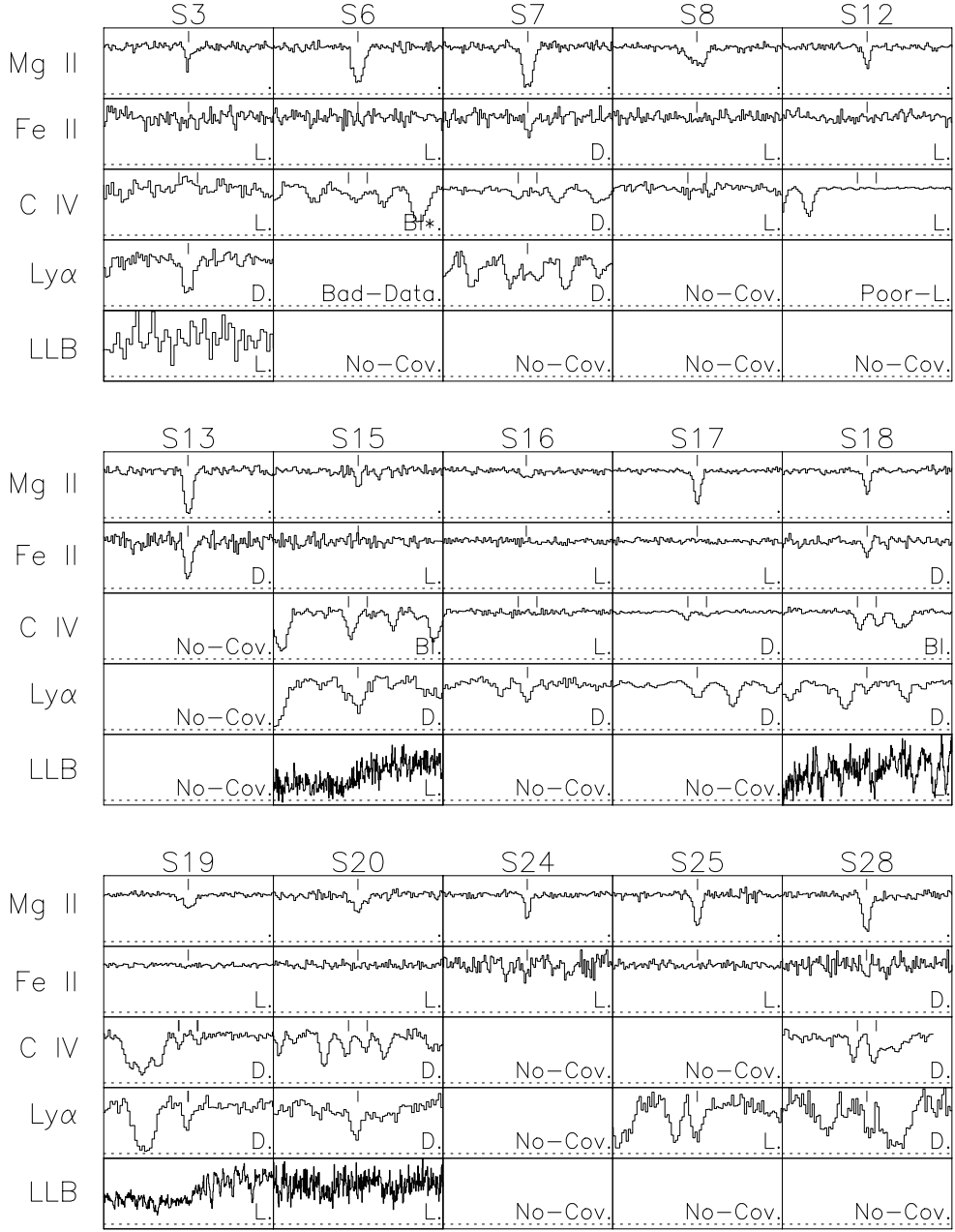}
\vglue -1.15in
\protect\caption{
The ``data matrix'' for the single--cloud weak {\MgII} systems.  For
each absorber the {\MgII} $\lambda 2796$ transition is shown in the
top subpanel.  In the respective lower subpanels are presented the
spectral regions where the {\FeII} $\lambda 2600$ (or $\lambda 2383$)
transition, the {\CIV} doublet, the {\Lya} transition, and the Lyman limit
break are expected.  Ticks above the spectra give the locations where
features are expected.  The full velocity window of the subpanels with
{\MgII} and {\FeII} is $100$~{\kms} and for the FOS data is $5000$~{\kms}.
``No--Cov'' indicates that the spectral region was not observed, and
``Bad--Data'' indicates that signal-to-noise ratio in the spectral
region was too low for a useful measurement.  ``D'' indicates a clean
detection at the $3\sigma$ or greater significance level.  ``L''
denotes no detection, but only an upper limit on the equivalent width.
``Bl'' indicates poor constraints due to blending with other features.
Transitions not plotted can be found in
Churchill {\etal} (2000a).
}
\label{fig:array}
\end{figure*}

\clearpage

\begin{figure*}
\figurenum{4}
\plotone{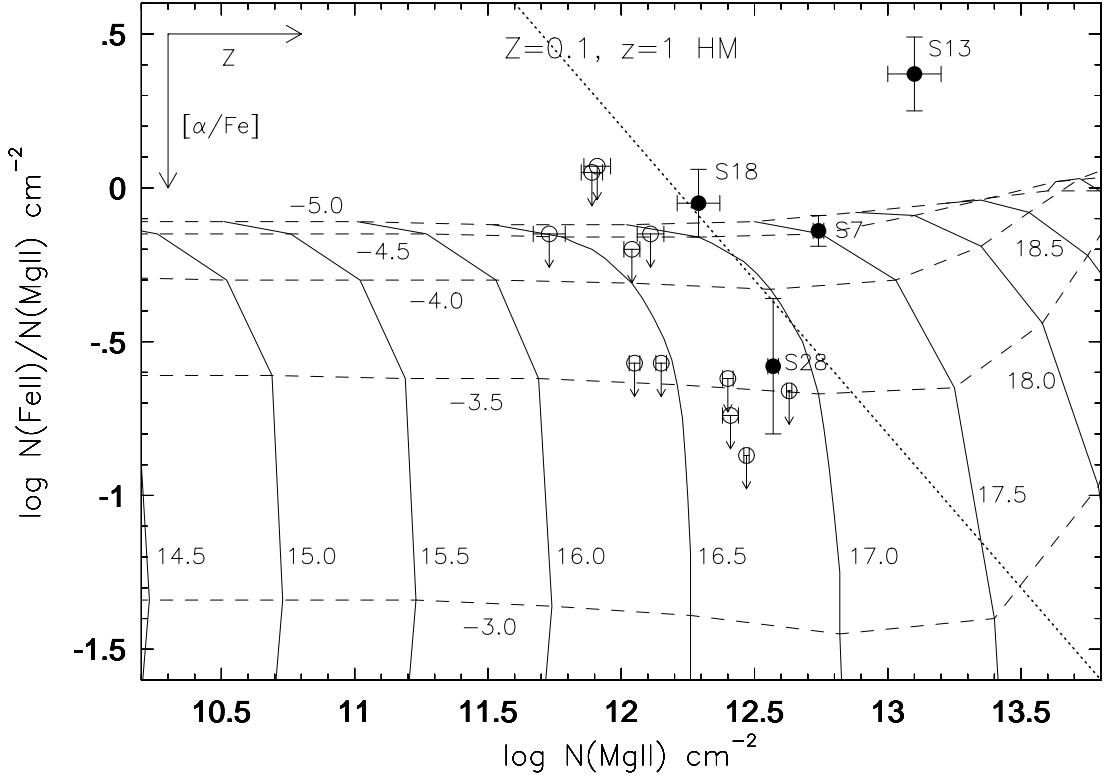}
\protect\caption{
The ratio of $N({\FeII})$ to $N({\MgII})$ versus $N({\MgII})$.  Clouds
with detected {\FeII} (filled circles) are identified by system
number.  Limits, depicted by open circles, are obtained from the $3\sigma$ 
equivalent width limits on {\FeII} $\lambda 2600$.  The dotted
diagonal line on the plot represents the $3\sigma$ detection limit
for {\FeII} in a spectrum with limiting equivalent width $W_{r}(2600) =
0.02$~{\AA}.  The individual system data points are superimposed on a
Cloudy grid for metallicity $Z=-1$ with a solar abundance pattern, and
using a Haardt--Madau (1996) spectrum at $z=1$.
Solid lines indicate constant $\log N({\HI})$ and dotted lines
indicate constant $\log U$.  Higher metallicity would shift the grid
to the right; $\alpha$--enhancement would shift the grid down.  In
order that S18 does not have a significant Lyman limit break, its
metallicity is constrained to be significantly larger than $Z=-1$.
High above the permitted grid of values, S13 may be iron--enhanced.
It is clear that a solar or slightly iron--enhanced abundance pattern
is required to produce the high $N({\FeII})/N({\MgII})$ ratios.
}
\label{fig:femg-grid}
\end{figure*}

\clearpage

\begin{figure*}
\figurenum{5}
\plotone{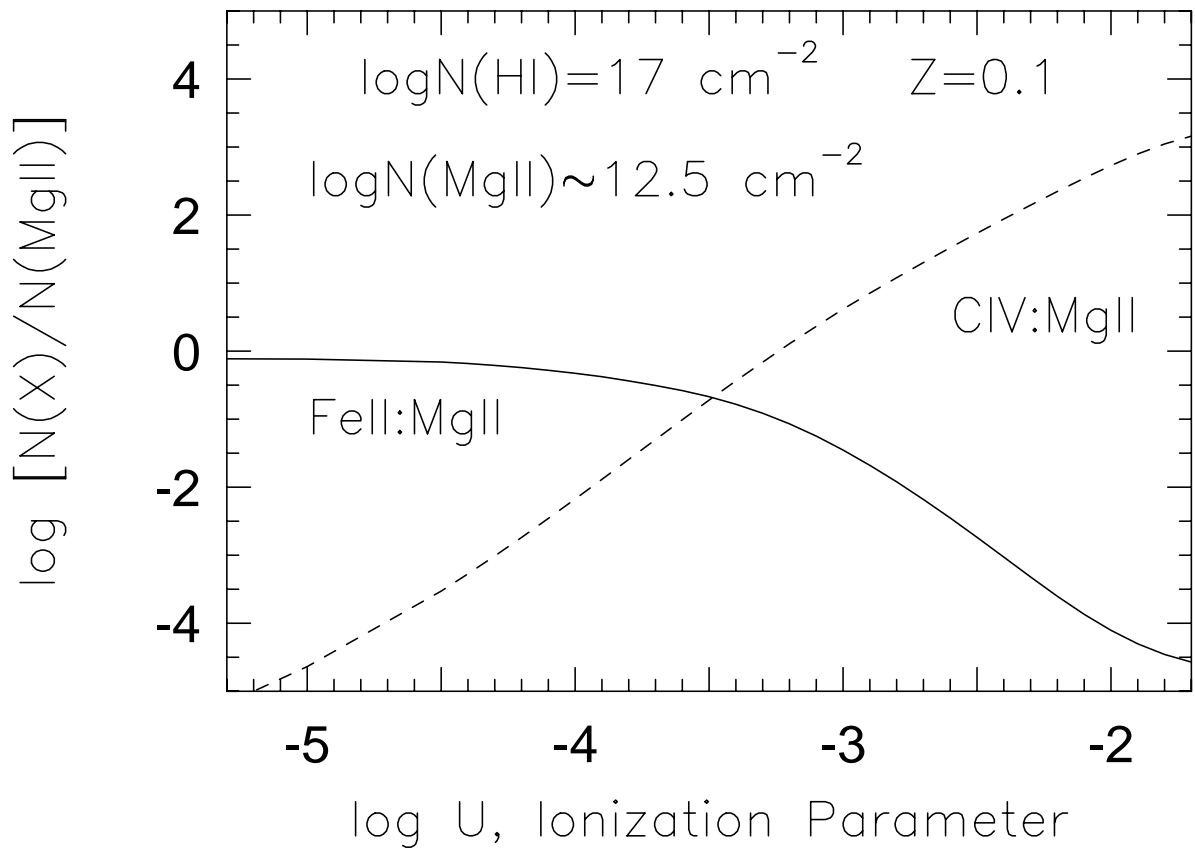}
\protect\caption{
The ratios $N({\FeII})/N({\MgII})$ and $N({\CIV})/N({\MgII})$ 
are uniquely determined functions of the ionization parameter over 
three dex of $\log U$.  Since ionization structure is not important for weak
{\MgII} absorbers, the ratios are 
independent of metallicity.  Note that at low values of $\log U$, the  
{\FeII}/{\MgII} ratio flattens and thus provides less constraint.
}
\label{fig:ratios}
\end{figure*}

\clearpage

\begin{figure*}
\figurenum{6}
\plotone{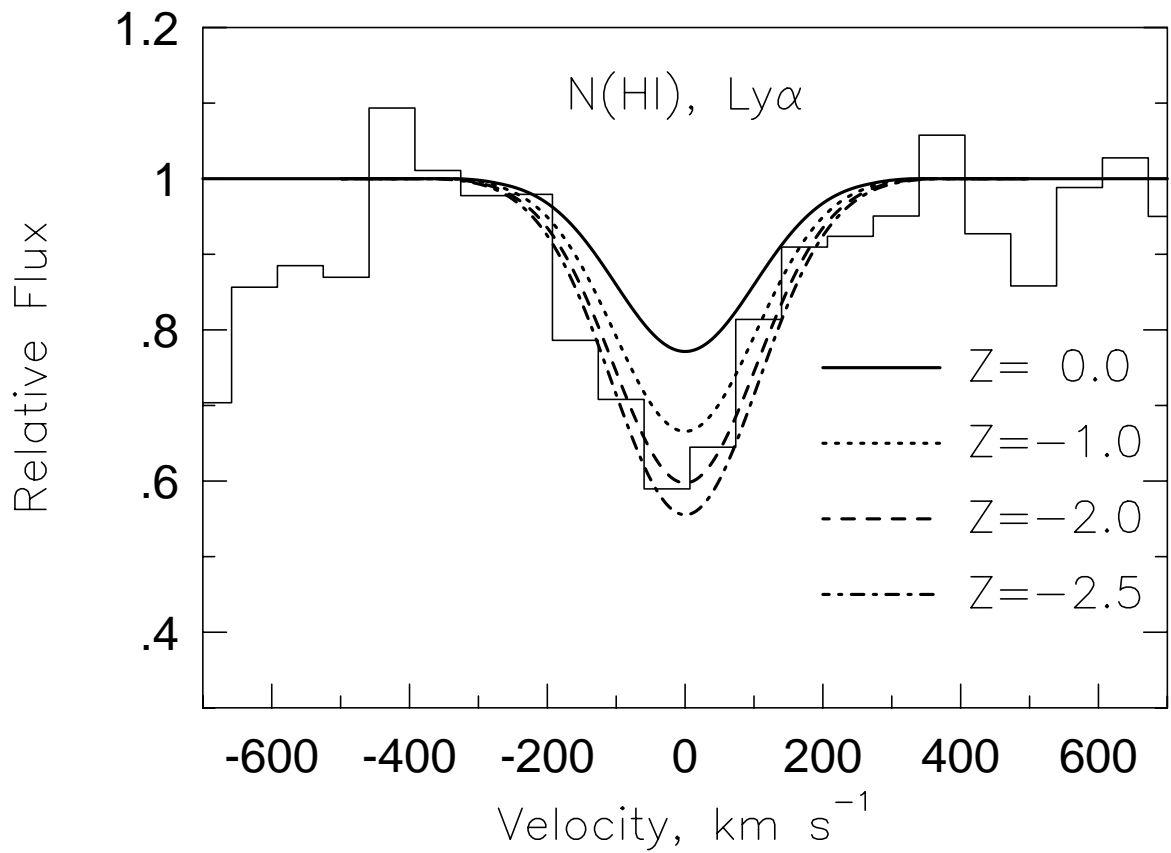}
\protect\caption{
Illustration of procedure to constrain cloud metallicity from {\Lya}.
Model predictions for metallicities of $Z=0$, $-1$, $-2$, and $-2.5$
are superimposed on the {\Lya} profile of S16.  Lower metallicities
predict more neutral hydrogen for a given observed N({\MgII}).
Clearly, $Z=0$ and $Z=-1.0$ do not fit the {\Lya} profile; thus, the
metallicity is constrained to be $-2.5<Z<-1.5$.  }
\label{fig:metdemo}
\end{figure*}

\clearpage

\begin{figure*}
\figurenum{7}
\plotone{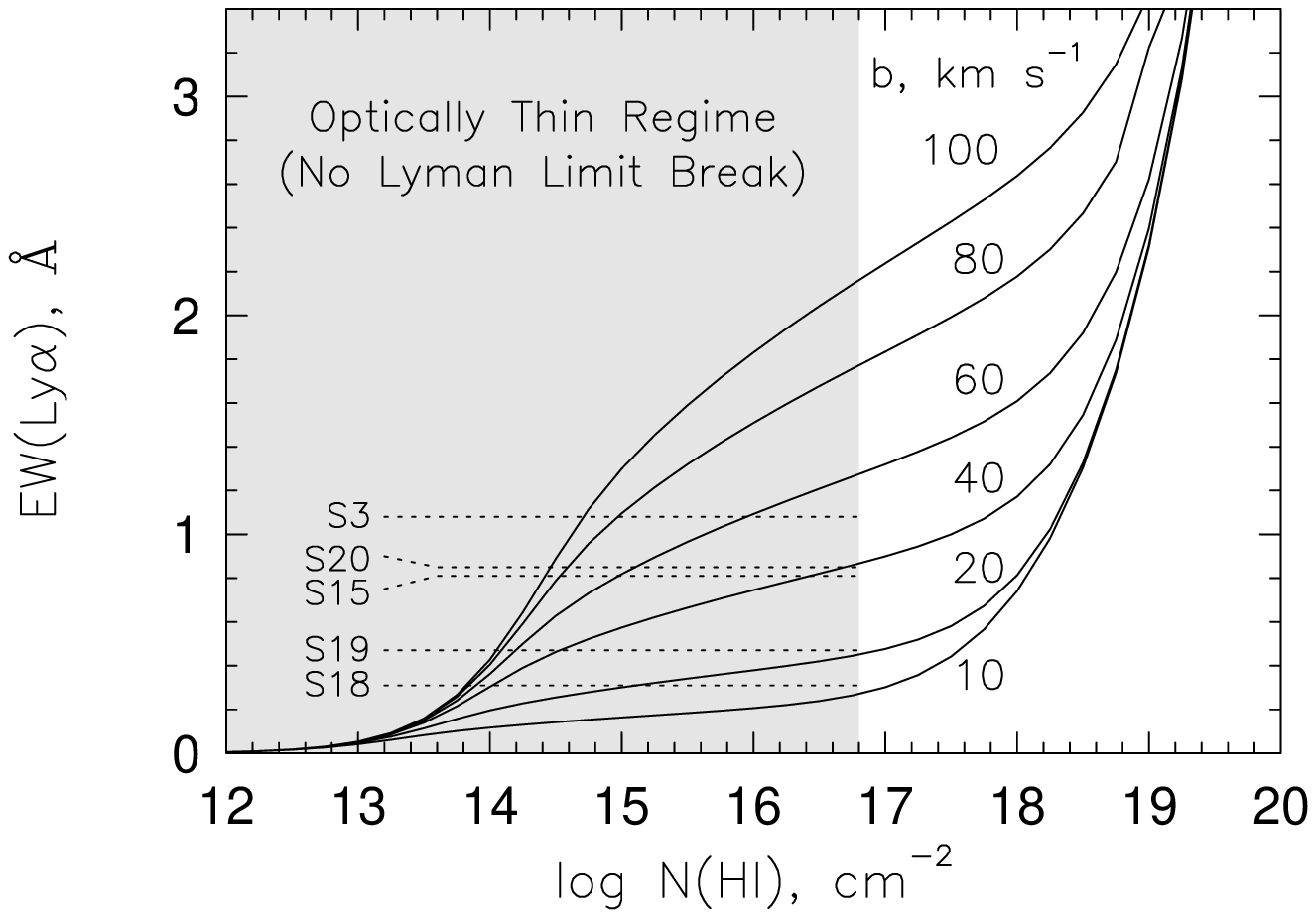}
\protect\caption{
Illustration of the constraints placed on the Doppler parameter, $b$,
of a broad {\Lya} component from the curve of growth of the {\Lya}
transition.  Curves of equivalent width versus {\HI} column density
are given for six different values of $b({\Lya})$, ranging from
$10$--$100$~{\kms}.  The equivalent widths of {\Lya} are shown as
horizontal lines for the five weak {\MgII} absorbers without a
detected Lyman limit break.  Restricting $\log N({\HI}) <
16.8$~{\cmsq} for the measured $W_{r}(2796)$ provides a lower limit on
$b({\Lya})$ from this curve of growth.  From these considerations,
systems S3, S15, and S20 have $b({\Lya}) \geq 40$~{\kms}.}
\label{fig:cog}
\end{figure*}

\clearpage

\begin{figure*}
\figurenum{8}
\plotone{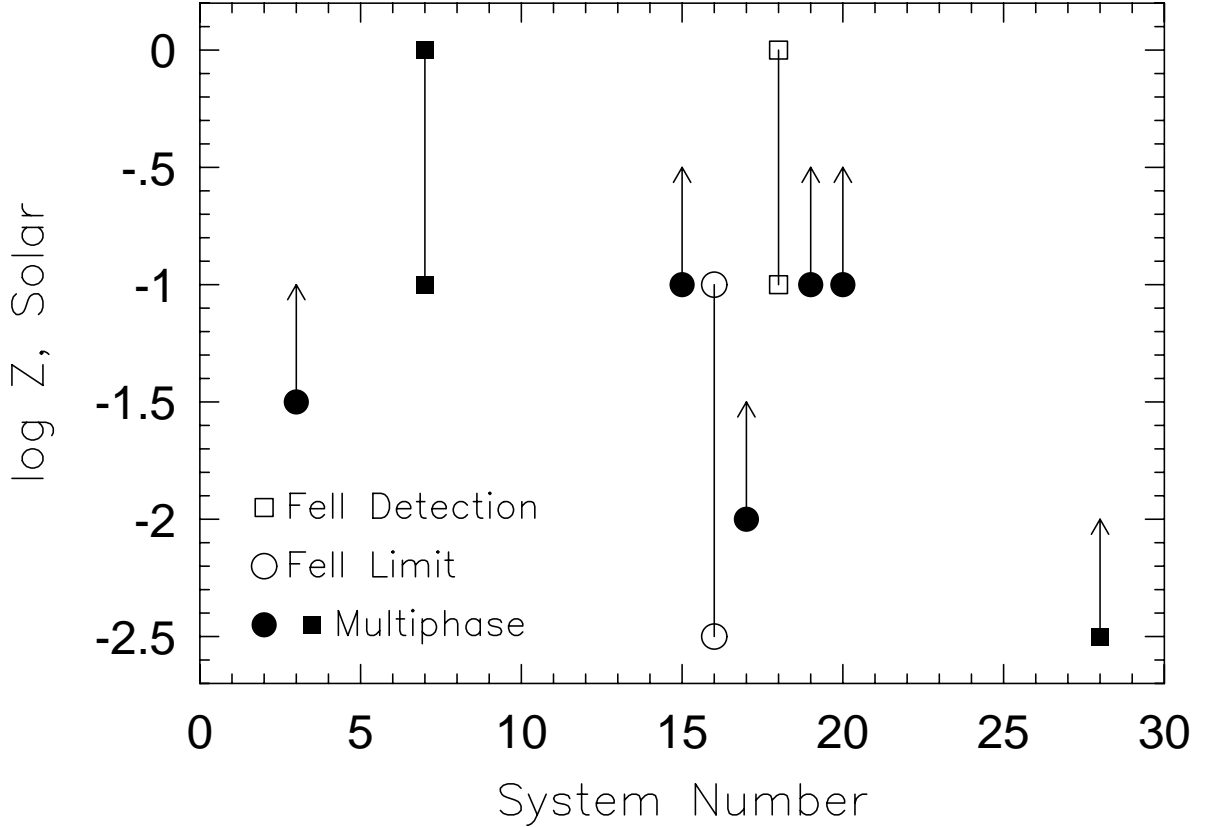}
\protect\caption{
Metallicities of single--cloud weak {\MgII} absorbers.  Metallicities
apply to the {\MgII} phase only, and are constrained fairly high.
Symbol type indicates whether {\FeII} was detected, and filled symbols
indicate that a second gas phase was also inferred.  Both {\Lya} and
the Lyman limit constrain Z.  The {\Lya} constraint is dependent
upon the assumed Doppler parameter, scaled from b({\MgII}).  By
contrast, lack of a Lyman limit break requires $\log N({\HI})<16.8$,
regardless of Doppler parameter.  When these constraints conflict, the
cloud has two phases of gas; the Lyman limit constraint is correct and
sets an upper limit on N({\HI}), whereas the {\Lya} constraint is in
error because much of the {\Lya} equivalent width arises in the second
phase.  The metallicity of the high--ionization phase cannot be
constrained with low--resolution spectra.  Metallicity only weakly
depends upon $U$; the quoted metallicity covers the range of permitted
$U$, which was constrained by {\CIV} limits and {\FeII} limits or
detections.  }
\label{fig:z}
\end{figure*}

\clearpage

\begin{figure*}
\figurenum{9}
\plotone{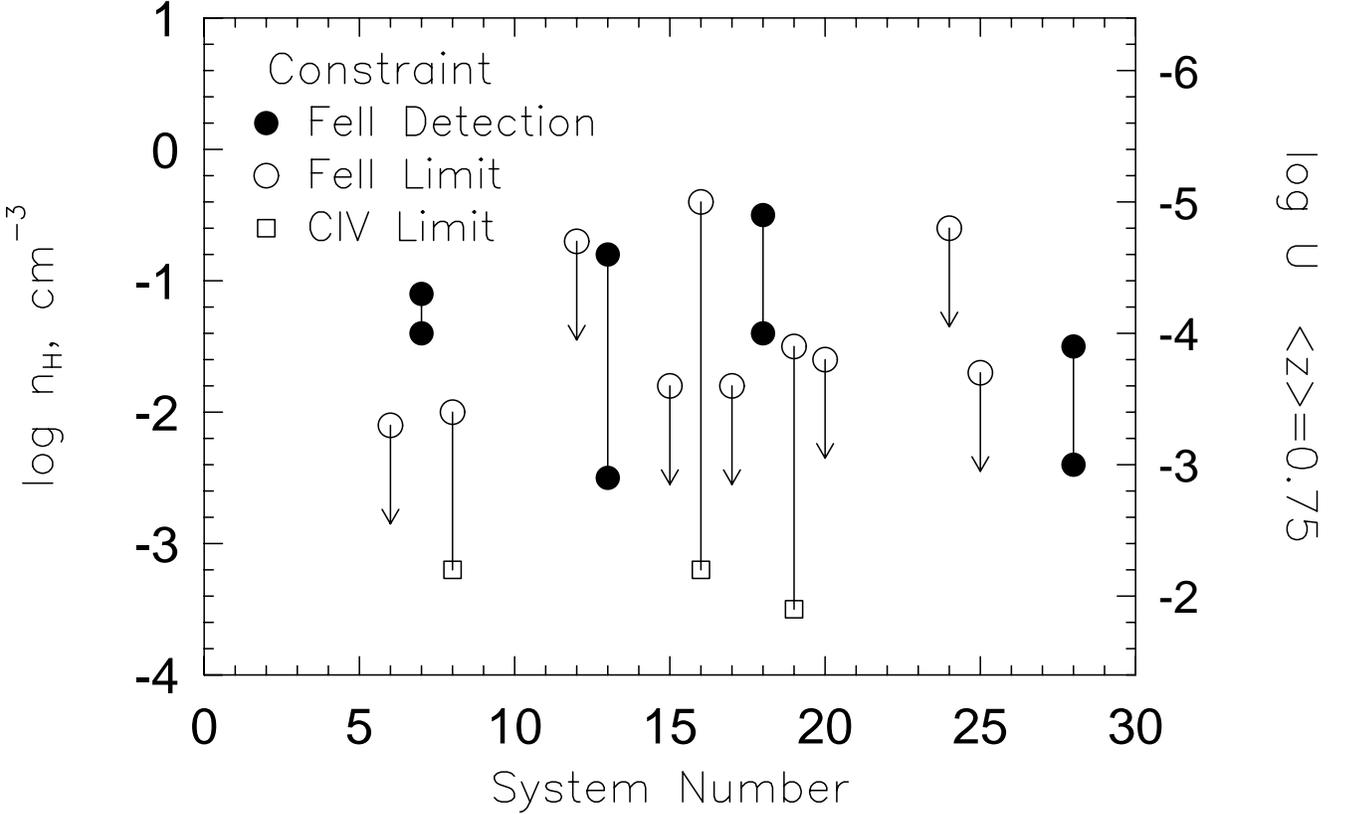}
\protect\caption{
Inferred Density and Ionization Parameter.  Ionization conditions were
determined in one of two ways: 1) by detection of {\FeII}, in which
case the vertical range between the two datapoints is the permitted
range of $\log U$ or 2) by {\FeII} limits and less frequently by
{\CIV} limits, which set lower and upper limits on $\log U$,
respectively.  Ionization parameter and density are related by
$U=n_{\gamma}/n_{\rm H}$, where n$_\gamma$, the number density of ionizing
photons, is redshift--dependent. For
display purposes, we have adopted $\log n_{\gamma} = -5.4$, an
intermediate value, to convert the derived densities into approximate
$\log U$.  To obtain the actual values of $\log U$ from the plot, add
$0.2$ dex to $\log U$ values for systems S8 and below, and subtract
$0.2$ dex to $\log U$ values for S12 and above.}
\label{fig:dens_ip}
\end{figure*}

\clearpage

\begin{figure*}
\figurenum{10}
\vglue -0.5in
\plotone{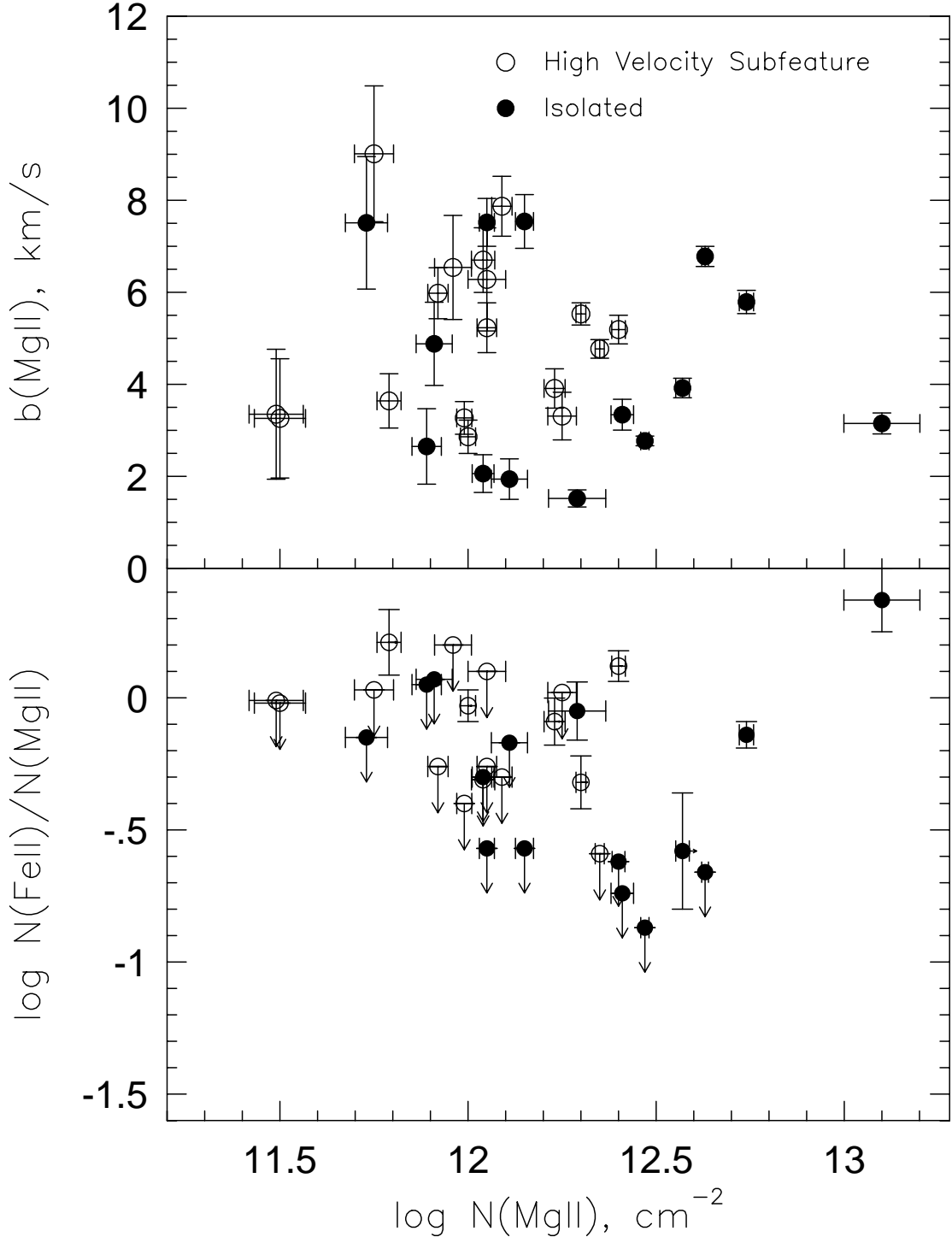}
\vglue -0.3in
\protect\caption{
Comparison of weak single--cloud {\MgII} absorbers and weak single--cloud 
outlying subfeatures ($50$--$400$~{\kms}) of strong {\MgII}
absorbers, for a limiting equivalent width of $W_r(2796)\le0.02$ {\AA}.  The
top panel plots the Doppler parameter $b({\MgII})$ versus $\log
N({\MgII})$, and shows little difference between the distributions for
the isolated and outlying clouds.  The lower panel shows the ratio of
$N({\FeII})$ to $N({\MgII})$, which is used to place constraints on
the ionization parameter, $\log U$.  Both the isolated and outlying
weak {\MgII} absorbers show a range of {\FeII} to {\MgII} ratios.  For
both types of systems, several objects are constrained to have small
thicknesses, $\sim 10$~pc.  }
\label{fig:outlying}
\end{figure*}

\end{document}